\documentclass[12pt,letterpaper]{article}
\usepackage{hyperref}
\usepackage{latexsym}
\usepackage{amssymb,amsfonts,amsmath}
\usepackage{graphicx}
\usepackage{indentfirst}
\usepackage{amsmath}
\usepackage{amsfonts}
\usepackage{amssymb}%
\usepackage{comment}
\ifx\pdfoutput\undefined
\else
\fi
\hypersetup{colorlinks=false,bookmarksopen,bookmarksnumbered,citecolor=blue,
pdfstartview=FitH}

\parskip=\medskipamount

\def\a{\alpha}

\def\b{\beta}

\def\d{\delta}
\def\e{\epsilon}

\def\g{\gamma}
\def\G{\Gamma}

\def\k{\kappa}
\def\l{\lambda}

\def\o{\omega}

\def\L{\Lambda}

\def\S{\Sigma}

\newcommand{\sect}[1]{\section{#1}\setcounter{equation}{0}}

\newcommand{\be}{\begin{equation}}
\newcommand{\ee}{\end{equation}}
\newcommand{\bea}{\begin{eqnarray}}
\newcommand{\eea}{\end{eqnarray}}

\newcommand{\ba}{\begin{array}}
\newcommand{\ea}{\end{array}}

\def\double #1{#1{\hbox{\kern-2pt $#1$}}}

\newcommand{\bsubeq}{\begin{subequations}}
\newcommand{\esubeq}{\end{subequations}}

 \newcommand{\virgolette}{``}

\begin{document}

\begin{titlepage}
\begin{flushright}
\par\end{flushright}
\vskip 1.5cm
\begin{center}
\textbf{\huge \bf Supersymmetric Wilson Loops \\ \vskip .3cm via Integral Forms}
\vskip 1cm
\vskip 0.5cm
\large {\bf C.~A.~Cremonini}$^{~a,b,}$\footnote{carlo.alberto.cremonini@gmail.com}, 
\large {\bf P.~A.~Grassi}$^{~c,d,e,}$\footnote{pietro.grassi@uniupo.it}, 
and
\large {\bf S.~Penati}$^{~f,}$\footnote{silvia.penati@mib.infn.it}, 
\vskip .5cm {
\small
\medskip
\centerline{$^{(a)}$ \it Dipartimento di Scienze e Alta Tecnologia (DiSAT),}
\centerline{\it Universit\`a degli Studi dell'Insubria, via Valleggio 11, 22100 Como, Italy}
\medskip
\centerline{$^{(b)}$ \it INFN, Sezione di Milano, via G.~Celoria 16, 20133 Milano, Italy} 
\medskip
\centerline{$^{(c)}$
\it Dipartimento di Scienze e Innovazione Tecnologica (DiSIT),} 
\centerline{\it Universit\`a del Piemonte Orientale, viale T.~Michel, 11, 15121 Alessandria, Italy}
\medskip
\centerline{$^{(d)}$
\it INFN, Sezione di Torino, via P.~Giuria 1, 10125 Torino, Italy}
\medskip
\centerline{$^{(e)}$
\it Arnold-Regge Center, via P.~Giuria 1,  10125 Torino, Italy}
\medskip
\centerline{$^{(f)}$ 
\it Dipartimento di Fisica, Universit\`a degli studi di Milano-Bicocca, }
\medskip
\centerline{\it and INFN, Sezione di Milano-Bicocca, Piazza della Scienza 3, 20126 Milano, Italy}
\medskip
}
\end{center}
\vskip  0.2cm
\begin{abstract}
We study supersymmetric Wilson loops from a geometrical perspective. To this end, we propose  a new formulation of these operators in terms of an integral form associated to the immersion of the loop into a supermanifold. This approach provides a unifying description of Wilson loops preserving different sets of supercharges, and clarifies the flow between them. Moreover, it allows to exploit the powerful techniques of super-differential calculus for investigating their symmetries. As remarkable examples, we discuss supersymmetry and kappa-symmetry invariance.

\end{abstract}
\vfill{}
\vspace{1.5cm}
\end{titlepage}
\newpage\setcounter{footnote}{0}
\tableofcontents

\sect{Introduction}

One of the most powerful techniques for studying general relativity is  the differential and Cartan calculus. 
Geometric objects like the Riemann tensor and its relatives are constructed in terms of {\em differential forms}
defined on a curved manifold. These intrinsic definitions require neither coordinates nor suitable 
parametrizations and are then independent of the specific model. In this 
formulation reparametrization invariance of the theory becomes manifest, which corresponds to the equivalence principle. 
Similarly, translating 
such a powerful technique into the framework of supersymmetric models promotes supersymmetry to a general super-repametrization invariance 
principle and supergravity arises as a natural consequence \cite{Witten:2012bg,Belo,Castellani:2014goa,Castellani:2015paa}. 

The basic ingredients of the differential supercalculus are \emph{integral forms} \cite{Witten:2012bg,Leites:1980rna,Manin,Deligne,Voronov}, which replace the notion of top forms in the context of a supermanifold, and whose integration over the entire supermanifold can be consistently defined. In particular, given an ordinary $p$-form $\omega^{(p|0)}$ on a supermanifold ${\cal SM}$ of dimensions $(n|m)$, its integration over a $p$-dimensional submanifold ${\cal N} \subset  {\cal SM}$ can be defined  as the integration on the entire supermanifold of the integral form $\omega^{(p|0)} \wedge \mathbb{Y}^{(n-p|m)}_{\cal N}$, where $\mathbb{Y}^{(n-p|m)}_{\cal N}$ is the \emph{Poincar\'e dual} to the immersion of ${\cal N}$ into ${\cal SM}$ \cite{Belo,Castellani:2015paa}. This is also named the {\em Picture Changing Operator} (PCO), being related to a similar concept in string theory (see e.g. \cite{Deligne,pol}).  

Integral forms have been already used to develop a geometric formulation of some simple topological theories such as 
super Chern-Simons theory \cite{GrassiMaccaferri,CremoniniGrassi,CremoniniGrassi2} towards $d=3$ $N=1$ supergravity. 
In this paper we apply this  technique to reformulate supersymmetric Wilson loops in terms of integral forms. We will primarily focus on $N=1$ SYM theory in ten dimensions and $N=4$ SYM in four dimensions, though the general technique that we propose can be adapted to theories in different dimensions and with different degrees of supersymmetry. 

The supersymmetric generalization of an ordinary Wilson loop \cite{Wilson:1974sk} appeared for the first time in \cite{Gates:1976rk}, for four dimensional gauge theories in $N=1$ superspace. Roughly speaking, it corresponds to replacing the ordinary path-ordered exponential as
\begin{equation}
P \, e^{\int_\l dx^\mu A_\mu } \quad \longrightarrow \quad P \, e^{\int_\L dz^M A_M }
\end{equation}
where $z^M = (x^a, \theta^\a, \bar{\theta}^{\dot \a})$ are superspace coordinates running on a supercontour $\L$ and $A_M = (A_a, A_\a, \bar{A}_{\dot \a})$ is the gauge superconnection. Further study of these operators has been done later in the development of supersymmetric field theories \cite{Gervais:1979zq, Marculescu:1980is, Mkrtchian:1982us, Awada:1995fe, Karp:2000mn}. More recently, this kind of operators have been investigated within the context of the AdS/CFT correspondence \cite{Drukker:1999zq, Ooguri:2000ps, Munkler:2015gja}, and integrability and Yangian invariance of the $N=4$ SYM theory \cite{Muller:2013rta, Beisert:2015jxa, Beisert:2015uda}. Light-like super-Wilson loops have been studied as dual to super-amplitudes in $N=4$ SYM \cite{CaronHuot:2010ek, Belitsky:2011zm, CaronHuot:2011ky, Beisert:2012xx}, also in a twistor formulation \cite{Mason:2010yk, Beisert:2012gb, Vergu:2016pzk}. 

Here we study an alternative geometric formulation, which makes use of integral forms. Our proposal is based on a preliminary observation regarding ordinary (bosonic) loops. 

For an ordinary Wilson operator $ W = {\rm Tr}_{\cal R} P\,e^{\Gamma}$ defined on a given $n$-dimensional manifold ${\cal M}$, the holonomy is given by $\Gamma = \int_\lambda A^{(1)}_*$, where $A^{(1)}_*$ is the pull-back of the gauge connection onto the path $\lambda$. Constructing the PCO ${\mathbb Y}_\lambda^{(n-1)}$ which describes the immersion of $\l$ in ${\cal M}$,  it can be rewritten as 
\begin{eqnarray}
\label{introB}
\Gamma = \int_{\cal M} A^{(1)} \wedge \mathbb{Y}^{(n-1)}_{\lambda}
\end{eqnarray}
Inside the integral  the connection spans the entire manifold, whereas all the information about the path is encoded in the PCO. The main advantage of this formulation is that $\G$ is manifestly invariant under diffeomorphisms of ${\cal M}$. 

This construction can be generalized to supersymmetric Wilson loops by promoting the top form appearing in the $\Gamma$ integral to
a supersymmetric top form. However, care is required to circumvent the well-known problem of defining a geometric 
formulation of integration on a supermanifold (not to be confused with Berezin integral on superspace). As already mentioned, this problem 
has been solved \cite{Leites:1980rna,Manin} and the central result is the substitution of the top form in $\Gamma$ with an integral form. 
Therefore, the geometric expression of a super-Wilson loop that we propose is the following 
\begin{eqnarray}
\label{introD}
{\cal W} =  {\rm Tr}_{\cal R} P \, e^{\Gamma} \qquad , \qquad \G = \int_{\cal SM} A^{(1|0)} \wedge \mathbb{Y}^{(n-1|m)}_{\L}
\end{eqnarray}
where $A^{(1|0)}$ is the gauge superconnection evaluated on the entire supermanifold ${\cal SM}$ and 
$\mathbb{Y}^{(n-1|m)}_{\L}$ is an integral form representing the Poincar\'e dual of the immersion of the 
supercontour  $\L$ into the supermanifold.  

In this formulation, invariance under superdiffeomorphisms is automatically implemented. Since $\G$ is factorized into the product of two objects, manifest invariance leads to the important identity 
\begin{eqnarray}
\label{introC}
{\delta} A^{(1)} \wedge \mathbb{Y}^{(n-1)}_{\lambda} + A^{(1)} \wedge {\delta}  \mathbb{Y}^{(n-1)}_{\lambda} \sim 0 \qquad {\rm up~to~}d-{\rm exact~terms}
\end{eqnarray}
which relates the variation of the Wilson loop to the variation of its supercontour, i.e. $(\d {\cal W})(\L) = - {\cal W}(\d \L)$.  It follows that the invariances of the Wilson operator are totally ascribable to the isometries of the PCO, which in turn can be investigated using differential geometry and the Cartan calculus. 

As we discuss in the main text, all the PCOs belong to the same $d$-cohomological class, i.e. the addition of a $d$-exact term does not change their defining properties. However, different representatives exhibit in general a different spectrum of isometries. This freedom of choosing a particular representative can be used to algebraically impose a given set of isometries on $\mathbb{Y}^{(n-1)}_{\L}$, leading to a Wilson loop that exhibits a given set of symmetries. 
We exploit this mechanism of {\em $d$-varying symmetries} to investigate the behavior of a super-Wilson loop under supersymmetry and kappa symmetry. A notable example is the BPS Wilson-Maldacena loop in $N=4$ SYM that we prove to be obtainable from the ordinary non-BPS operator by 
the addition of a suitable $d$-exact term to the original PCO. 

It is remarkable to note that our formulation of Wilson operators can be easily generalized to the case of curved (super)manifolds, so leading to Wilson operators in (super)gravity, which is technically built-in.

The paper is structured as follows. In section 2 we briefly review the main tools of superdifferential calculus, integral forms, Poincar\'e duals and the geometric construction of super Yang-Mills theories. Section 3 is focused on the geometrical construction of abelian Wilson loops along the lines described above, both for the bosonic and the supersymmetric cases. Within the present geometric framework, in section 4 we investigate Wilson loop invariance under a reparametrization of the path, superdiffeomorphisms, supersymmetry and kappa symmetry. In particular, we show how Killing spinor equations corresponding to BPS Wilson loops arise in the present formalism. In section 5 the generalization to Wilson loops in non-abelian gauge theories is briefly presented. Finally, section 6 contains a brief discussion about the interesting relation between our geometric construction of Wilson operators and a similar construction in the context of pure spinor string theory. A brief summary of our main results and a discussion on possible follows-up can be found in section 7. Three appendices follow, which provide some technical material to supplement the main text and the equations therein.

\sect{Preliminaries}

In this section we briefly review the geometric construction of supersymmetric gauge theories in terms of superdifferential forms. We mainly restrict to definitions and properties that will be useful in the following sections. For a more extensive introduction to this topic we refer the reader for example to \cite{Witten:2012bg,Belo,Castellani:2014goa,Castellani:2015paa,CremoniniGrassi,Noja}.

\subsection{Superdifferential Forms}

Superdifferential forms, and more specifically integral forms, are the crucial ingredients that allow
to define a geometric integration theory for supermanifolds \cite{Witten:2012bg} inheriting all the good properties of 
differential form integration theory in conventional geometry.  

We recall that in the space of differential superforms there is no notion of top form, that is a form that can be suitably integrated on the supermanifold. This is due to the commuting nature of the fundamental one-forms $d\theta$'s corresponding to odd $\theta$-coordinates. As proposed in \cite{Leites:1980rna, Manin, Deligne}, the notion of top form has to be found into a new complex of forms known as integral forms. Here we follow the strategy pioneered by Belopolsky \cite{Belo}, where
integral forms are distributional-like forms on which a suitable Cartan calculus can be developed. We clarify the basic ingredients and rules, and refer to the literature \cite{Witten:2012bg, CremoniniGrassi, CatenacciGrassiNoja, CatenacciGrassiNoja2} for a complete description.

We consider a supermanifold ${\cal SM}^{(n|m)}$ with $n$ bosonic and $m$ fermionic dimensions. We denote the local coordinates in an open set as $(x^a, \theta^\alpha)$. 
A $(p|q)$-form 
$\omega^{(p|q)}$ has the following structure
\begin{eqnarray}
\label{genA}
\omega^{(p|q)} (x, \theta, dx, d\theta)  = \omega(x,\theta) \, dx^{a_1}\dots dx^{a_r} \, d\theta^{\a_1} \dots d\theta^{\a_s} \, \delta^{(b_1)}
(d\theta^{\beta_{1}}) \dots  \delta^{(b_{q})}(d\theta^{\beta_{q}})
\end{eqnarray}
where the $d\theta^\a$ appearing in the product are independent of those 
appearing in the deltas ($\a_i \neq \b_j$ for any pair $i,j$) and the $b_i$ indices denote the 
number of derivatives acting on  the delta functions. The $\omega(x,\theta)$ coefficients, explicitly given by $\omega_{[a_1 \dots a_r](\a_1 \dots \a_s)[\beta_1 \dots \beta_q]}(x,\theta)$, are a set of superfields.  The indices $a_1 \dots a_r$ and $\beta_1 \dots \beta_q$ are anti-symmetrized, whereas the 
indices $\alpha_1 \dots \alpha_s$ are symmetrized, because of the rules\footnote{We also recall the following properties (the $\alpha$ index is not summed)
\begin{eqnarray}
\label{genB}
d \, \delta^{(a)} \! (d\theta^\a) = 0\,, ~~~~~~
d\theta^\a \delta^{(a)} \!(d\theta^\a) = - a \delta^{(a-1)} \! (d\theta^\a)\,, ~~ a >0\,, ~~~~~
d\theta^\a \delta (d\theta^\a) = 0 \nonumber
\end{eqnarray} }
\begin{eqnarray}
\label{genC}
&&dx^a dx^b = - dx^b dx^a\,, ~~~
dx^a d\theta^\a = d\theta^\a dx^a\,, ~~~
d\theta^\a d\theta^\b = d\theta^\b d\theta^\a\,, \\
&&
\delta(d\theta^\a) \delta(d\theta^\beta)  = - \delta(d\theta^\beta)  \delta(d\theta^\alpha) \,, ~~~
dx^a \delta(d\theta^\a) = - \delta(d\theta^\alpha) dx^a\,, ~~~
d\theta^\a \delta(d\theta^\b) = \delta(d\theta^\b) d\theta^\a  \nonumber
\end{eqnarray}
From the first identity of the second line we note that $\delta ( d \theta )$ has to be treated as an anticommuting object, unlike the standard $\delta$ distribution. This is due to the fact that $\delta ( d \theta )$ is used to compute the \emph{oriented volume} of the supermanifold. Indeed, $\delta \left( d \theta \right)$ is not a distribution on smooth functions, but rather on \virgolette smooth differential forms". This is mathematically called a \emph{de Rham current} (see \cite{Witten:2012bg} for further explanations).

The two quantum numbers $p$ and $q$ in eq. \eqref{genA} correspond to the {\it form} number and the {\it picture} number, respectively, and they range as $-\infty < p < +\infty$ 
and $0 \leq q \leq m$. The total form degree is given by $p = r + s - \sum_{i=1}^{i=q} b_i$
since the derivatives act effectively as negative forms and the delta functions do not carry any form degree.  The total picture $q$ of $\omega^{(p|q)}$ corresponds 
to the number of delta functions. In particular, we call it {\it superform} if $q=0$, 
\begin{eqnarray}
\label{superform}
\omega^{(p|0)} (x, \theta, dx, d\theta)  = \omega(x,\theta) dx^{a_1}\dots dx^{a_r} d\theta^{\a_1} \dots d\theta^{\a_s},    \qquad p = r+s \ ,
\end{eqnarray}
or {\it integral form} if $q=m$, 
\begin{eqnarray}
\label{integralform}
\omega^{(p|m)} (x, \theta, dx, d\theta)  = \omega(x,\theta) dx^{a_1}\dots dx^{a_r}  \delta^{(b_1)}
(d\theta^{\beta_{1}}) \dots  \delta^{(b_{q})}(d\theta^{\beta_{q}}) \ , \ p = r - \sum_{i=1}^q b_i
\end{eqnarray}
Otherwise it is called {\it pseudoform}.   

A top integral form $\omega^{(n|m)}$ corresponds to an element of the line bundle known as Berezinian bundle (the transition functions are represented 
by the superdeterminant of the Jacobian) and it can be locally expressed as in eq. \eqref{integralform} with $p=n$.  
 As in  conventional geometry, we can define the integral of a top form on the superspace $T^* {\cal SM}$ 
endowed with a super-measure $[dx d\theta d(dx) d(d\theta)]$ as
 \begin{eqnarray}
\label{genD}
I[\omega] = \int_{{\cal SM}} \omega^{(n|m)} = 
\int_{T^* {\cal SM}} \omega(x, \theta, dx, d\theta) [dx d\theta d(dx) d(d\theta)]
\end{eqnarray}
where the order of the integration variables is kept fixed and the measure is invariant under coordinate transformations. We refer the reader to \cite{Witten:2012bg} for a complete discussion on the symbol $[dx d\theta d(dx) d(d\theta)]$. Here we simply recall that while $dx$ and $d(d\theta)$ are ordinary Lebesgue integrals, the integrations over $d\theta$ and 
$d(dx)$ are Berezin integrals. Therefore, the following identities hold
\begin{eqnarray}
\int dx \, d[dx] \equiv \int \delta({dx}) \, d[dx] = 1 \qquad , \qquad \int \delta({d\theta}) \, d[d\theta] = 1
\end{eqnarray}
where in the first relation we emphasised the fact that being $dx$ an odd variable, it coincides with its Dirac delta function. Performing the Berezin $d[dx]$ integrations and the algebraic $d[d \theta]$ ones in \eqref{genD}, it is then easy to check that $I[\omega]$ is nothing but the ordinary superspace integral  
\begin{eqnarray}
I[\omega] = \int_{{\cal SM}} \omega(x, \theta) dx^1 \dots dx^n d\theta^1 \dots d\theta^m  
\end{eqnarray}
of the $\omega(x, \theta)$ superfield. In the present formulation the Stokes theorem for integral forms is also valid. 

By changing the one-forms $dx^a, d\theta^\a$ as 
\begin{equation}\label{change}
dx^a \rightarrow E^a = E^a_m dx^m + E^a_\mu d\theta^\mu \qquad , \qquad d\theta^\a \rightarrow E^\a = E^\a_m dx^m + E^\a_\mu d\theta^\mu
\end{equation}
a top form $\omega^{(n|m)}$  transforms as
\begin{eqnarray}
\label{genEAB}
\omega^{(n|m)} \rightarrow {\rm Ber}( E)  \, \omega(x,\theta) \, dx^1 \dots dx^n \, \delta(d\theta^1) \dots \delta(d\theta^m)
\end{eqnarray}
where $ {\rm Ber}(E)$ is the superdeterminant (i.e. \emph{Berezinian}) of the supervielbein $(E^a, E^\a)$.

\subsection{Picture Changing Operators (PCO)}
 
The strategy that we use for constructing suitable integral forms to be integrated on a supermanifold is the following.
Given a bosonic $p$-form $\omega^{(p|0)}$, which can be integrated on a $p$-dimensional submanifold ${\cal N} \subset {\cal SM}$, 
we note that since its integral is not on the entire supermanifold, its
transformation properties under superdiffeomorphisms are not manifest. They become manifest if the integral can be converted into the integral of an integral form over the entire ${\cal SM}$. Indeed, this can be achieved by constructing the Poincar\'e dual form 
${\mathbb Y}^{(n-p|m)}_{\cal N}$ of the immersion\footnote{Precisely, we consider 
 ${\cal N} \subset {\cal M}$ where ${\cal M}$ is the bosonic component of  $ {\cal SM}$ known in the literature as the body.} of ${\cal N}$ into ${\cal SM}$ (a.k.a. {\it Picture Changing Operator} (PCO)) such that, if we denote $\omega_*^{(p|0)} \equiv \iota_* \omega^{(p|0)}$ with $\iota$ the  immersion of ${\cal N}$ into ${\cal SM}$, we define
\begin{eqnarray}
\label{introA}
 \int_{\cal N} \iota_*\omega^{(p|0)} = \int_{\cal SM} \omega^{(p|0)} \wedge \mathbb{Y}^{(n-p|m)}_{\cal N}
\end{eqnarray}
The second expression is the integral over the whole supermanifold of a $(n|m)$-dimensional top form to which we can apply the usual Cartan calculus rules.  

This is a well-known formula 
in differential geometry (see for example Bott and Tu \cite{botttu}) which allows us to disentangle the geometrical 
properties of the immersed surfaced ${\cal N}$ in the entire manifold or supermanifold from the properties of the $\omega^{(p|0)}$
integrand. In topological field theories it is a powerful tool used to prove the Duistermaat-Heckman formula \cite{DH} for the localization technique 
and to implement the computations in that framework using the Thom isomorphism \cite{cordes}. 

The PCO in \eqref{introA} is independent of the coordinates, it only depends on the immersion 
through its homology class. Moreover, it is closed, but not exact
\begin{equation}\label{genEAC}
	d{\mathbb Y}^{(n-p|m)}=0 \ , \qquad  \mathbb{Y}^{(n-p|m)} \neq d \Sigma^{(n-p-1|m)}
\end{equation}
 Therefore, by changing the immersion $\iota$ to an homologically equivalent surface ${\cal N}'$, the new Poincar\'e dual 
 $\mathbb{Y}^{(n-p|m)}_{\cal N'}$  differs from the original one by $d$-exact terms.   
 It is important to note that if $\omega^{(p|0)}$ is a closed form, then  (\ref{introA}) is automatically invariant under any change of the embedding (we will always assume there are no boundary contributions). 
 
Tipically, for rigid supersymmetric models the closed form $\omega^{(n|0)}$ is represented by 
the Lagrangian of the model ${\cal L}^{(n|0)}(\Phi,V, \psi)$ built using the rheonomic rules (see \cite{cube}). It 
is a function of dynamical superfields $\Phi$ and the rigid supervielbeins $V^a, \, \psi^\a$ defined in eq. \eqref{supervielbeins}. The corresponding action reads
\begin{eqnarray}
S = \int_{{\cal SM}^{(n|m)}} {\cal L}^{(n|0)}(\Phi, V, \psi) \wedge  {\mathbb Y}^{(0|m)}(V,\psi)
\end{eqnarray} 
where the PCO ${\mathbb Y}^{(0|m)}$ contains only geometric data (for instance the supervielbeins or the coordinates 
themselves). If  $d  {\cal L}^{(n|0)}(\Phi, V, \psi)  =0$ we can change the PCO by exact terms without changing the action. This can be conveniently exploited for choosing for instance a PCO that possesses manifest symmetries. 

In the supergravity case, after the change \eqref{change}, $(E^a, E^\alpha)$ are promoted to dynamical fields and the action becomes 
\begin{eqnarray}
\label{genG}
S_{sugra} = \int_{{\cal SM}^{(n|m)}} {\cal L}^{(n|0)}(\Phi, E) \wedge  {\mathbb Y}^{(0|m)}(E)
\end{eqnarray}
The closure of the lagrangian and the closure of the PCO imply the conventional supergravity constraints that
reduce the spectrum of independent fields to the one of physical fields.

\subsection{Geometry of Supersymmetric Abelian Gauge Fields}

We recall some basic facts about the geometrical construction of supersymmetric gauge theories. We focus first on the $U(1)$ case, postponing the review of the non-abelian case to section \ref{sect:NAGT}.
As a starting point we consider the 10d N=1 SYM case, since other cases can be obtained by dimensional reduction 
and suitable truncations.

The 10d gauge supermultiplet is represented by one vector superfield and one spinor superfield (the gaugino) with 
degrees of freedom matching on-shell \cite{Nilsson:1981kx,Witten:1985nt,Harnad:1985bc}. No superspace off-shell formulation is known, which includes the correct spectrum of auxiliary fields allowing to 
construct a superspace action that leads to the correct equations of motion. 
However, a super-geometric formulation can be developped, which stems from promoting the gauge field to a superfield
$(1|0)$-superform  (with $V^a$ and $\psi^\a$ defined in \eqref{supervielbeins})
\begin{eqnarray} \label{superconnection}
A^{(1|0)} = A_a V^a + A_\a \psi^\a
\end{eqnarray}
The corresponding field strength, defined as  
\begin{eqnarray}\label{fieldstrength}
F^{(2|0)} \equiv d A^{(1|0)} = F_{ab} V^a V^b + F_{a\alpha} V^a \psi^\a + F_{\a\b} \psi^\a \psi^\b
\end{eqnarray}
is subject to Bianchi identities supplemented by the conventional gauge invariant constraint 
\begin{eqnarray}
F_{\a\b} \equiv D_{(\a} A_{\b)} + \gamma^a_{\a\b} A_a =0 
\end{eqnarray}
from which we obtain $A_a$ as a function of the spinorial components $A_\a$. As a consequence, the other components turn out to be uniquely expressed 
in terms of the gaugino $(0|0)$-superform $W^\a$ 
\begin{eqnarray}
\label{SYMC}
F_{a\a}  =   (\gamma_a W)_\a \qquad , \qquad F_{ab}  =    (D \gamma_{ab}W) 
\end{eqnarray}
and satisfy the additional constraints
\begin{eqnarray}\label{SYMC2}
D^\a W_\a =0 \qquad , \qquad D_\a F_{ab} = (\gamma_{[a} \partial_{b]}  W)_\a 
 \end{eqnarray}  
These constraints automatically imply the equations of motion for all the physical fields. 

By using suitable gamma matrices identities, one can prove that the previous relations can be recast in the following superform equations 
\begin{eqnarray}
\label{SYMB}
F^{(2|0)} = V^a\wedge V^b F_{ab} + (\psi \gamma_a W) V^a \qquad , \qquad
d W^\a = V^a \partial_a W^\a - \frac14 (\gamma^{ab} \psi)^\a F_{ab} 
\end{eqnarray}  

The great advantage of using the geometric formulation is that supersymmetry transformations can be expressed as superdiffeomorphisms along the fermionic directions (see appendix \ref{app1} for the geometric definition of supersymmetry transformations). In particular, the gauge superfields transform as
\begin{eqnarray}
\label{SYMD}
\delta_\epsilon A^{(1|0)} &=& {\cal L}_\epsilon A^{(1|0)} = \iota_\epsilon dA^{(1|0)} - d (\iota_\epsilon A^{(1|0)}) \nonumber \\
&=& \epsilon \gamma_a W V^a+ 4 \epsilon \gamma^a \theta V^b F_{ab} - 2 \epsilon \gamma^a \theta \psi \gamma_a W  -  d (\iota_\epsilon A^{(1|0)}) \nonumber \\
\delta_\epsilon W &=& {\cal L}_\epsilon W = \iota_\epsilon dW = - \frac14 (\gamma^{ab} \epsilon) F_{ab}
\end{eqnarray}
These relations give rise to the ordinary supersymmetry transformations up to a gauge transformation of the gauge field $A$, while the gaugino superfield $W^\a$ is  gauge covariant. We note that these rules remain true also in the case of local transformations.

\subsection{Dimensional Reduction}\label{sect: dimred}

As is well--known, $D=4 , N=4$ SYM theory can be obtained by dimensional reduction of the $D=10, N=1$ theory, while preserving the maximal amount of supersymmetry. Here we clarify how to perform the dimensional reduction in the geometric set-up. 

Given the set of ten dimensional superspace coordinates $(x^a, \theta^\a)$, $a=0, \dots , 9$ and $\a =1 , \dots , 16$, we decompose $x^a = (x^{\a \dot{\a}}, y^{[AB]})$ and $\theta^\a = (\theta^{A \a}, \bar{\theta}_A^{\dot{\a}})$, where $\a, \dot\a = 1,2$ are spinorial indices in Weyl representation and $A = 1, \dots, 4$ are $SU(4)$ R-symmetry indices.

Starting from the ten dimensional superform \eqref{superconnection}, we first perform the following decompositions  
\begin{eqnarray}
\label{decoA}
A_a V^a = A_{\a\dot \a} V^{\a\dot\a} + \phi_{[AB]} V^{[AB]}\,, ~~~~~ 
A_\a \psi^\a = A_{A,\a} \psi^{A, \a} + \bar A_{\dot\a}^A  \, \bar{\psi}^{\dot{\a}}_A
\end{eqnarray}
Here $V^{\a\dot\a}$ can be identified with the components of the four-dimensional vielbein, whereas $V^{[AB]}$ is the vielbein along the extra six directions. It satisfies the self-duality constraint $\bar V_{AB} = \epsilon_{ABCD} V^{[CD]}$. Similarly, $\psi = (\psi^{A,\a}, \bar{\psi}_A^{\dot{\a}})$ represents the decomposition of the rigid gravitino field. They satisfy the following equations  
\begin{eqnarray}
\label{decoB}
&&d V^{\a\dot\a} = \bar\psi^{\dot\a}_A \psi^{A, \alpha}\,, ~~~~~
d V^{[AB]} = \epsilon^{ABCD} \epsilon_{\dot\a\dot\b}  \bar\psi^{\dot\a}_C  \bar\psi^{\dot\beta}_D + 
\epsilon_{\a\b} \psi^{A, \alpha} \psi^{B, \beta}  ~~~~~ \nonumber \\
&&d \psi^{A, \a} =0 \,, ~~~~~ d \bar\psi^{\dot\a}_A =0 
 \end{eqnarray}
 
In the same way, we decompose the gaugino superform  $W_\a = (W^A_{\a}, \bar W_{A \, \dot\a})$ according to its $SL(2,\mathbb{C}) \times SU(4)$ representation. 
 
The dimensional reduction is then achieved by removing the dependence of the fields upon the 
transverse coordinates $y^{[AB]}$. The four $A_{\a\dot{\a}}$ components then describe the gauge connection in four dimensions, $\phi_{[AB]}$ are the six real scalars of the $N=4$ SYM theory and $W_{A, \a}$ give rise to the four gaugini. As a consequence, from the definition of the field strength $F^{(2|0)} $ in (\ref{SYMB}) we obtain
\begin{eqnarray}
\label{decoCA}
F^{(2|0)} &=& 
V^{\a\dot \a}\wedge V^{\b\dot\b} F_{{\a\dot \a}, {\b\dot\b}} + 
2\, V^{\a\dot \a} \wedge V^{AB} F_{{\a\dot \a}, AB} + 
 \nonumber \\
&- &  V^{\a\dot \a}  (\bar\psi_{A, \dot\a} W^A_\a + \psi^A_\alpha \bar W_{A, \dot \a}) - 
 V^{AB} (\bar\psi_{A, \dot\a} \bar W_B^{\dot \a} +  \epsilon_{ABCD} \psi^{C}_\a W^{D,\a})
\end{eqnarray}
As described in \cite{DAuria:1981zjr}, in order to complete the dimensional reduction we have to redefine the connection as 
\begin{eqnarray}
\label{decoD}
A^{(1|0)} \, \rightarrow \, A^{(1|0)} - \Phi_{[AB]} V^{[AB]} 
\end{eqnarray}
where $\Phi_{[AB]}$ are six chiral superfields containing the $\phi_{[AB]}$ scalars. As a consequence, the superfield strength becomes 
\begin{eqnarray}
\label{decoC}
\hspace{-0.1cm} F^{(2|0)} = 
V^{\a\dot \a}\! \wedge \! V^{\b\dot\b} F_{{\a\dot \a}, {\b\dot\b}} \! - \! V^{\a\dot \a} 
(\bar\psi_{A, \dot\a} W^A_\a  +  \psi^A_\alpha \bar W_{A, \dot \a}) 
 \!+  \! (\epsilon^{ABCD} \epsilon_{\dot\a\dot\b}  \bar\psi^{\dot\a}_C  \bar\psi^{\dot\a}_D + 
\epsilon_{\a\b} \psi^{A, \alpha} \psi^{B, \beta})  \Phi_{AB}  \nonumber \\
\end{eqnarray}
and coincides with the expression for the superfield strength of the $N=4$ SYM theory obtained directly in four dimensional non-chiral superspace (see for instance \cite{cube}). 
We note that additional pieces proportional to the flat gravitinos appear, which carry an explicit dependence on the scalar fields $\phi_{[AB]}$. As we are going to explain in the next sections, these terms are crucial for the construction of the supersymmetric version of BPS Wilson loops in four dimensions. 

\subsection{Geometry of Supersymmetric Non-Abelian Gauge Fields}\label{sect:NAGT}

We now review the geometric construction for non-abelian gauge fields, still focusing on the ten dimensional case \cite{Nilsson:1981kx,Witten:1985nt,Harnad:1985bc}. As for the abelian case, a superspace off-shell formulation of gauge superfields with auxiliary fields is not known, but a geometric formulation can be provided. 

For a non-abelian gauge group the superfield strength is defined as 
\begin{eqnarray}\label{NAFT}
F^{(2|0)} \equiv d A^{(1|0)} + \frac12 A^{(1|0)} \wedge A^{(1|0)} = 
F_{ab} V^a V^b + F_{a\alpha} V^a \psi^\a + F_{\a\b} \psi^\a \psi^\b
\end{eqnarray}
and it is subject to the Bianchi identities
\begin{eqnarray} 
\nabla F^{(2|0)} =0
\end{eqnarray}
with the covariant derivative defined as $ \nabla F^{(2|0)}  = d F^{(2|0)} + [ A^{(1|0)}, F^{(2|0)}] =0$. 
This is supplemented by the conventional gauge invariant constraint 
\begin{eqnarray}
F_{\a\b} \equiv \nabla_{(\a} A_{\b)} + \gamma^a_{\a\b} A_a =0 
\end{eqnarray}
from which one obtains $A_a$ as a function of the spinorial components $A_\a$. The other components turn out to be expressed in terms of the gaugino superfield $W^\a$ as
\begin{eqnarray} \label{SYMCn}
F_{a\a}  =   (\gamma_a W)_\a\,, \qquad 
W^\alpha = \gamma^{a \alpha \beta} (\nabla_a A_\beta - \nabla_\beta A_a)\,, 
\qquad F_{ab}  =    \nabla^\a (\gamma_{ab})_{\a \b} W^\b 
\end{eqnarray}
and satisfy the additional constraints
\begin{eqnarray}\label{SYMC2n}
\nabla^\a W_\a =0\,,  \qquad \nabla_\a F_{ab} = (\gamma_{[a} \nabla_{b]}  W)_\a 
 \end{eqnarray}  
Equations (\ref{SYMCn}), (\ref{SYMC2n}) imply the equations of motion, which are then a consequence of the superspace constraints. 

Supersymmetry transformations are easily expressed as 
\begin{eqnarray}
\label{SYMDn}
\delta_\epsilon A^{(1|0)} &=& {\cal L}_\epsilon A^{(1|0)} =
 \iota_\epsilon \left( dA^{(1|0)}  + \frac12 A^{(1|0)}\wedge A^{(1|0)} \right) + 
 d (\iota_\epsilon A^{(1|0)}) + [A^{(1|0)},  \iota_\epsilon A^{(1|0)}] \nonumber \\
&=& \epsilon \gamma_a W V^a+ 4 \epsilon \gamma^a \theta V^b F_{ab} - 2 \epsilon \gamma^a \theta \psi \gamma_a W  + \nabla (\iota_\epsilon A^{(1|0)}) \nonumber \\
\delta_\epsilon W &=& {\cal L}_\epsilon W = \iota_\epsilon \nabla W - \left[ \iota_\epsilon A^{(1|0)} , W \right] = - \frac14 (\gamma^{ab} \epsilon) F_{ab} - \left[ \iota_\epsilon A^{(1|0)} , W \right]
\end{eqnarray}
These relations give rise to the ordinary supersymmetry transformations up to a gauge transformation of the gauge field $A$, while the gaugino superfield $W^\a$ is  gauge covariant. 

The reduction to four dimensions works exactly as for the abelian case, section \ref{sect: dimred}, with the obvious covariantization of the equations.

\sect{Geometric construction of a supersymmetric Wilson loop: The abelian case}\label{sect:abelian}

We present a general construction of supersymmetric Wilson loops in terms of integral forms. 
The main goal is to obtain a general expression suitable for any geometry of the loop and whose invariances are easily analysable. 
For the time being we restrict to the case of an abelian gauge theory. The generalization to the non-abelian case is discussed in section \ref{nonabelian}. 

\subsection{Ordinary Wilson Loops as Integral Forms}\label{sect:bosonicWL}

As a warm-up, we begin by discussing how to write ordinary (i.e. bosonic) Wilson loops in terms of integral forms. 

Given an abelian gauge theory with gauge connection $A^{(1)}$ defined on a manifold ${\cal M}$ of arbitrary dimension $n$, a Wilson loop along a curve $\lambda \subset {\cal M}$ is given by
\begin{eqnarray}
\label{WLA}
W = e^{\Gamma}  \qquad ,\qquad \Gamma =\int_\lambda A^{(1)}_*
\end{eqnarray}
where $A^{(1)}_* $ is the pull-back of the connection one-form $A^{(1)} = A_a dx^a$ along the curve. The integration of a one-form ensures the parametrization independence of the loop. As usual, by choosing a suitable parametrization, one can compute the integral. 

When $\lambda$ is a closed path the $W$ operator is gauge invariant. This can be made manifest by alternatively expressing the Wilson loop in terms of the curvature two-form $F^{(2)} = d A^{(1)}$. In fact, using the Stokes theorem we can rewrite $\Gamma$ as an integral over a two dimensional surface ${\cal S}$ whose boundary is $\lambda$
\begin{eqnarray}
\label{WLB}
\Gamma =\oint_\lambda A^{(1)}_* = \int_{\cal S} F^{(2)}
\end{eqnarray}
This expression is then manifestly invariant under gauge transformations. 

We now prove that $\Gamma$ can be rewritten as the integral of an $n$-form on the entire manifold ${\cal M}$. To this end we introduce the PCO dual 
to the immersion of the one-dimensional curve $\lambda$ into the manifold ${\cal M}$  
\begin{eqnarray}
\label{WLC}
{\mathbb Y}^{(n-1)}_\lambda  = \prod_{i=1}^{n-1} \delta(\phi_i) \delta (d\phi_i) \equiv  \prod_{i=1}^{n-1} \delta(\phi_i) \, d\phi_i
\end{eqnarray}
where $\{ \phi_i \}_{i=1, \dots , n-1} $ is a set of $(n-1)$ functions whose zero locus
\begin{eqnarray}
\label{WLD}
\lambda = \{ x \in {\cal M} |~~ \phi_i(x) = 0\,, ~~~~ i = 1\,, \dots\,, n-1\} 
\end{eqnarray}
defines the curve $\lambda \subset {\cal M}$. In the second equality we have used $\delta(d\phi_i) = d\phi_i$, being $d\phi_i$ anticommuting differential one-forms.

As a simple example we consider the unit circle in two dimensions. In this case we have a single function $\phi(x_0, x_1) = x_0^2 + x_1^2 - 1$ whose locus defines the curve. The Poincar\'e dual to the immersion is then 
\begin{eqnarray}
\label{WLCAB}
{\mathbb Y}^{(1)} = 2 \delta( x_0^2 + x_1^2 - 1)(x_0 dx_0 + x_1 dx_1)
\end{eqnarray}
and it is manifestly invariant under the $O(2)$ isometry group of the circle\footnote{For the 2d manifold ${\cal M}$ where the 
circle is immersed, we can use the invariant vielbeins $V_{ang} = x_0 dx_1 - x_1 dx_0, V_{rad}= x_0 dx_0 + x_1 dx_1$, 
which are the usual angular and radial vielbeins $V_{ang} = d\phi$ and $V_r = r dr$.}.

The PCO in \eqref{WLC} possesses the following fundamental properties 
\begin{eqnarray}
\label{WLE}
d {\mathbb Y}^{(n-1)}_\lambda  &=& 0, \qquad  {\mathbb Y}^{(n-1)}_\lambda  \neq  d \eta^{n-2} \nonumber \\
\delta_\phi {\mathbb Y}^{(n-1)}_\lambda  &=& 
 d \left[  \prod_i \delta(\phi_i) (\sum_j \delta \phi_j \iota_j) \delta(d \phi_i) \right]
\end{eqnarray}
where $\iota_j$ is the contraction along the vector field $\partial_j$ and acts as $\iota_j \delta(d\phi_i)= \partial /\partial (d\phi_j) \delta(d\phi_i)$, while $\delta \phi_j$ is the variation of the constraints.  

The first identity can be proven by using the 
chain rule $d \delta(\phi_i) = (\sum_j d\phi_j \frac{\partial}{\partial \phi_j}) \delta(\phi_i)$  (the differential $d\phi_j$ is kept on the left hand side of the delta) and the distributional property $d\phi_i \delta(d\phi_i) =0$. To prove the second identity one needs to list all possible candidates for $\eta^{(n-2)}$ and then check that there is none. 
The last identity is more elaborated and makes use of the additional distributional identity (integration by parts) $d\phi_i \iota_i \delta(d\phi_i) = - \delta(d\phi_i)$ ($i$ is not summed) \cite{Belo}. 
In particular, it states that any variation of ${\mathbb Y}^{(n-1)}_\lambda $ by changing the immersion of the curve $\lambda$ into ${\cal M}$ 
is $d$-exact. In other words, each homologically equivalent curve $\lambda$ corresponds to a single cohomological class represented by ${\mathbb Y}^{(n-1)}_\lambda $.

Given a path $\lambda$ in ${\cal M}$ and the corresponding Poincar\'e dual ${\mathbb Y}^{(n-1)}_\lambda $ as in \eqref{WLC} the Wilson loop holonomy \eqref{WLA} can be rewritten in the following way
\begin{eqnarray}
\label{WLF}
\Gamma  =\int_\lambda A^{(1)}_* =   \int_{\cal M} A^{(1)}\wedge {\mathbb Y}^{(n-1)}_\lambda 
\end{eqnarray}
that is as a top form integrated over the entire manifold. The two expressions are clearly equivalent, but their interpretation is rather different. On the 
left hand side,  the connection is computed on a submanifold corresponding to the curve suitably parametrized. On the
right hand side instead, the connection is a generically assigned abelian gauge field on ${\cal M}$ while the geometrical data concerning the path are entirely captured by the PCO. In particular, the latter can be modified as ${\mathbb Y}^{(n-1)}_\lambda  \to {\mathbb Y}^{(n-1)}_\lambda  + d\S^{(n-2)}$, while preserving properties (\ref{WLE}) and leaving the connection unchanged. This freedom can be exploited to enhance the set of manifest symmetries of $\G$; these algebraic properties embody the strength of this method, since it would be much more difficult to ascribe these properties to the curve $\lambda$, namely on the homology side.

The $\Gamma$ integral in \eqref{WLF} is manifestly invariant under gauge transformations and deformations of the path within the class of homologically equivalent contours. 

Gauge invariance is manifest thanks to the closure property of Poincar\'e duals (first equation in \eqref{WLE}). In fact, under a gauge transformation $\delta A^{(1)} = d \a$ the integral transforms as
\begin{eqnarray}
\label{WLG}
\d \Gamma = \int_{\cal M} d \a \wedge {\mathbb Y}^{(n-1)}_\lambda = \int_{\cal M}  d \big( \a \wedge {\mathbb Y}^{(n-1)}_\lambda \big)  
\end{eqnarray} 
and the r.h.s. vanishes if $\partial {\cal M} =\emptyset$ or if we impose $\a$ to vanish at the intersection $\l \cap \partial {\cal M} $.

Invariance of the Wilson loop under a deformation of the path is also easy to study. In fact, from the last identity in \eqref{WLE} it turns out that a deformation of the path equations amounts to a shift of ${\mathbb Y}^{(n-1)}_\lambda$ by an exact term $d \eta^{(n-2)}$. Therefore, integrating by parts, we have 
\begin{eqnarray}
\label{AKAC}
\delta_\phi  \int_{\cal M} A^{(1)}\wedge {\mathbb Y}^{(n-1)}_\lambda =  
\int_{\cal M} A^{(1)}\wedge \delta_\phi {\mathbb Y}^{(n-1)}_\lambda = \int_{\cal M} F^{(2)} \wedge \eta^{(n-2)}
\end{eqnarray}
and the r.h.s. vanishes if the connection has zero curvature on the surface connecting the loop and its deformation, namely if the curve $\lambda$ has been deformed without encountering 
singularities. This shows the equivalence between Wilson loops computed on homologically equivalent curves. 

It is interesting to investigate how to recast in this new framework the identity in \eqref{WLB} which states the equivalence between the line integral of the connection $A^{(1)}$ and the surface integral of the field strength $F^{(2)}$. 
Given a surface ${\cal S}$ with $\partial {\cal S} = \lambda$, we call ${\mathbb Y}^{(n-2)}_{\cal S}$ the PCO dual to the surface immersed in the space ${\cal M}$. Therefore, we can write
\begin{eqnarray}
\label{RelA-F}
\int_{\cal S} F^{(2)}_*=\int_{\cal M} F^{(2)} \wedge {\mathbb Y}_S^{(n-2)} = \int_{\cal M} A^{(1)} \wedge d {\mathbb Y}_S^{(n-2)}  
\end{eqnarray}
where we have assumed that $d$-exact terms integrate to zero. As discussed above, Stokes theorem (or equivalently eq. \eqref{WLB}) implies  
$ d {\mathbb Y}^{(n-2)}_{\cal S} = {\mathbb Y}^{(n-1)}_\lambda$,
where ${\mathbb Y}^{(n-1)}_\lambda $ is the PCO of the path $\lambda$. However, this condition 
seems to violate the second identity in \eqref{WLE}. 

This apparent contradiction can be sorted out by observing that ${\mathbb Y}^{(n-2)}_{\cal S}$ does not have compact support, 
while $ {\mathbb Y}^{(n-1)}_\lambda $ is a distribution with compact support. 
In order to elaborate on this point we assume that locally we can split the manifold as ${\cal M} = {\cal M}' \times {\mathbb R}^+$, with 
the factor ${\mathbb R}^+$ described by the additional coordinate $x'$.
We take $\lambda$ to be immersed into ${\cal M}'$ only and the surface ${\cal S}$ to be
the union  ${\cal S} = \lambda \cup \{x' >0\}$. Moreover, we denote by 
${\mathbb Y}^{(n-2)}_{\lambda\subset {\cal M}'}$ the PCO dual of $\lambda$ in ${\cal M}'$ while
${\mathbb Y}^{(n-1)}_\lambda$ is still the dual of $\lambda$ in ${\cal M}$. 
If we define 
\begin{eqnarray}
\label{RelA-FB}
{\mathbb Y}^{(n-2)}_{\cal S} = \Theta(x') {\mathbb Y}^{(n-2)}_{\lambda \subset {\cal M}'}
\end{eqnarray}
where $\Theta(x')$ is the Heaviside theta function, a non-compact support distribution 
equal to $1$ for $x'>0$, it follows that
\begin{eqnarray}
\label{RelA-FC}
\hspace{-0.5cm} d {\mathbb Y}^{(n-2)}_{\cal S} = d \left(  \Theta(x') {\mathbb Y}^{(n-2)}_{\lambda \subset {\cal M}'} \right) = 
 d  \Theta(x') \wedge  \left(  {\mathbb Y}^{(n-2)}_{\lambda \subset {\cal M}'} \right) = 
 \delta(x') dx' \wedge  \left(  {\mathbb Y}^{(n-2)}_{\lambda \subset {\cal M}'} \right) = 
  {\mathbb Y}^{(n-1)}_\lambda
\end{eqnarray}
This is the expected identity which establishes relation \eqref{WLB} in the language of integral forms. 

Before closing this section, we give a simple formula for the bosonic 
Wilson loop and the corresponding PCO when the curve is parametrized as $\tau \rightarrow x^a(\tau)$, with $\tau \in {\mathbb T}  \subseteq {\mathbb R}$. 

We enlarge the manifold to ${\cal M} \times \mathbb{T}$ with coordinates $(x^\mu, \tau)$ and we construct the 
PCO dual to the embedding $\tau \rightarrow (x^\mu(\tau), \tau)$ as follows
\begin{eqnarray}
\label{AKACA}
{\mathbb Y}^{(n)}_\l &=& \prod_{a=1}^n \delta\Big(x^a - x^a(\tau) \Big) \bigwedge_{a=1}^n (dx^a - \dot{x}^a d\tau)  \nonumber \\
 &=&\prod_{a=1}^n \delta\Big(x^a - x^a(\tau) \Big)  
 \left( \bigwedge_{a=1}^n dx^a  + \sum_{b=1}^n (-1)^b  \dot x^b d\tau \bigwedge_{a\neq b} dx^a \right) 
\end{eqnarray} 
It then follows that
\begin{eqnarray}
\label{AKACB}
A^{(1)} \wedge {\mathbb Y}^{(n)}_\l 
 &=&  A_c dx^c \wedge \prod_{a=1}^n
 \delta\Big(x^a - x^a(\tau) \Big)  
 \left(\sum_{b=1}^n (-1)^b  \dot x^b d\tau \bigwedge_{a\neq b} dx^a \right)  \nonumber \\
 &=& A_c \dot x^c  d\tau \, \prod_{a=1}^n  \delta\Big(x^a - x^a(\tau) \Big)  \bigwedge_{a=1}^n dx^a 
\end{eqnarray}
where $\dot x^c = \frac{d x}{d\tau}^c$. Integrating on ${\cal M}\times \mathbb{T}$ we obtain
\begin{eqnarray}
\label{AKACC}
\int_{{\cal M}\times  \mathbb{T}} A^{(1)} \wedge {\mathbb Y}^{(n)}_\l  = 
\int_\lambda  d\tau  \, \dot x^c  (\tau)   A_c(x(\tau))  
\end{eqnarray}
which is the usual expression for a Wilson loop along $\lambda$ parametrized by $\tau$. 

To summarise, we have proposed a new expression for the holonomy of a bosonic Wilson operator as the integral of a top form on the entire manifold (see eq. \eqref{WLF}). To our knowledge this is a new formulation, which has never appeared in the literature before. It has the advantage to split the field and the contour dependences, making the investigation of invariances easier. Moreover, it allows for a natural generalization to the supersymmetric case, as we are going to discuss in the next section.

\subsection{Supersymmetric Wilson loops as Integral Forms}\label{sect:susyWL}

The supersymmetric version of eq. \eqref{WLA} can be defined as \cite{Gates:1976rk,GGRS}  
\begin{eqnarray}\label{susyW}
{\cal W} = e^{\Gamma} \qquad ,\qquad \Gamma =\int_\Lambda A^{(1|0)}_*
\end{eqnarray}
where $A^{(1|0)}_*$ is the pull-back of the connection superform on a supercurve $\Lambda$ defined in a supermanifold ${\cal SM}$ and parametrized by a set of local  coordinates $z^M(\tau) = (x^a(\tau), \theta^\a(\tau))$, $a=1, \dots, n$ and $\a = 1, \dots , m$. For example, in ten dimensional $N=1$ superspace ($n=10, \, m=16$) the connection superform is given by \eqref{superconnection}, and using definitions \eqref{supervielbeins} it can be explicitly written as  
\begin{eqnarray}\label{susyW2}
A^{(1|0)}_* = \left[  A_a  (\dot{x}^a + \theta \gamma^a \dot{\theta} )+  A_\a \dot{\theta}^\a\right ] d\tau 
\end{eqnarray}
Similarly, in four dimensional $N=1$ superspace ($n=m=4$), the corresponding gauge superform reads
\begin{eqnarray}
A^{(1|0)}_* = \left[  A_a  (\dot{x}^a + \theta \gamma^a \dot{\bar \theta} + \bar\theta \gamma^a \dot{\theta} )+  A_\a \dot{\theta}^\a + \bar{A}_{\dot{\a}} \dot{\bar \theta}^\a \right ] d\tau \qquad \a =\dot{\a} = 1,2 
\end{eqnarray}

For closed supercontours, ${\cal W}$ in \eqref{susyW}  is a non-local operator, invariant under supergauge transformations $\d A^{(1)} = d \omega$. Its lowest component coincides with the ordinary Wilson loop in \eqref{WLA}. 

Generalizing the procedure used in the bosonic case, we construct a super-Poincar\'e dual which localizes the integrand on the supercurve and allows to rewrite $\Gamma$ in \eqref{susyW} as an integral over the entire supermanifold. Precisely, if the immersion equations of the supercurve $\Lambda$ in ${\cal SM}$ are
\begin{eqnarray} \label{AKADgeneral}
\phi_a(x, \theta) &=& 0\, \quad a = 1, \dots , n-1 \nonumber \\
g^\a(x, \theta) &=& 0 \, \quad \a = 1, \dots , m
\end{eqnarray}
with $\{\phi_a\}$ a set of bosonic superfields in ${\cal SM}$ and $\{ g^\a \}$  a set of fermionic ones,
we introduce a factorized PCO ${\mathbb Y}^{(n-1|m)}_\Lambda \equiv  {\mathbb Y}^{(n-1|0)}_\Lambda \wedge {\mathbb Y}^{(0|m)}_\Lambda$, with 
\begin{eqnarray}
\label{AKAEgeneral}
\nonumber {\mathbb Y}^{(n-1|0)}_\Lambda  &=& \prod_{a=1}^{n-1} \delta(\phi_a(x,\theta)) \delta (d\phi_a) = \prod_{a=1}^{n-1} \delta(\phi_a(x,\theta)) d\phi_a  \\
{\mathbb Y}^{(0|m)}_\Lambda &=& \prod_{\alpha=1}^m \delta( g^\a(x, \theta)) \delta(dg^\a) = \prod_{\alpha=1}^m g^\a(x, \theta) \delta(dg^\a)
\end{eqnarray}
The second PCO carries no form degree, but it carries picture number equal to $m$. 

Assigned the PCO, we can rewrite the Wilson loop exponent $\G$ in \eqref{susyW} as
\begin{eqnarray}
\label{AKAF}
\Gamma = \int_{\cal SM} A^{(1|0)} \wedge {\mathbb Y}^{(n-1|m)}_\Lambda 
\end{eqnarray}
The superconnection is generically defined on ${\cal SM}$, while the geometrical data featuring the supercurve are captured by the Poincar\'e dual ${\mathbb Y}^{(n-1|m)}_\Lambda$. 

This  expression for $\Gamma$ can be made more explicit if we parametrize the supercurve $\Lambda$ in terms of smooth functions $ \tau \to z^M(\tau)$ on ${\mathbb T} \subseteq {\mathbb R}$. For the bosonic part of the PCO we can proceed exactly as done in section \ref{sect:bosonicWL} by including $\tau$ as an extra bosonic coordinate and extending the integration to the supermanifold ${\cal SM} \times {\mathbb T}$. A straightforward supersymmetrization of eq. \eqref{AKACA} leads to 
\begin{eqnarray}\label{BKA2}
{\mathbb Y}^{(n|0)}_\Lambda = \prod_{a=1}^n \delta\Big(x^a - x^a(\tau) \Big) \bigwedge_{a=1}^n (V^a - \Pi^a(\tau) d\tau) 
\end{eqnarray}
where we have defined $V^a(\tau) \equiv \Pi^a (\tau) d\tau= (\dot{x}^a +  \theta \gamma^a \dot{\theta})d\tau$.

For the PCO of the fermionic sector we choose
\begin{eqnarray}
\label{BKA}
{\mathbb Y}^{(0|m)}_\Lambda &=& \prod_{\alpha=1}^m (\theta^\a - \theta^\a(\tau)) \delta\Big(\psi^\a - \dot \theta^\a(\tau) d\tau\Big) \\
&=& \prod_{\alpha=1}^m (\theta^\a - \theta^\a(\tau)) 
\Big(1 - \sum_\beta \dot \theta^\b(\tau) d\tau \iota_\b\Big) \prod_{\alpha=1}^m \delta(\psi^\a) \nonumber
\end{eqnarray}
where in the second line we have expanded the Dirac delta functions exploiting the presence of the anticommuting one-form $d\tau$. Here $ \iota_\b$ is the contraction along the $D_\beta$ vector field. Using a shorter notation we can then write
\begin{eqnarray}\label{BKA3}
{\mathbb Y}^{(n|m)}_\Lambda &\equiv&  {\mathbb Y}^{(n|0)}_\Lambda \wedge {\mathbb Y}^{(0|m)}_\Lambda  = \\
&=& \d^{(n)} (x - x(\tau)) \, ( V - \Pi(\tau) d\tau)^n \, \wedge \, (\theta - \theta(\tau) )^m \, \d^{(m)}(\psi - \dot{\theta} (\tau) d\tau ) \nonumber
\end{eqnarray}
Focusing on the fermionic part, we can write $\G$ as
\begin{eqnarray}
\label{BKB}
&& \hspace{-0.8cm} \Gamma = 
\int_{{\cal SM} \times {\mathbb T}} A^{(1|0)}  \wedge {\mathbb Y}^{(n|m)}_\Lambda \\
&& \hspace{-0.5cm} =\int_{{\cal SM} \times {\mathbb T}} A^{(1|0)} \wedge 
\prod_{\alpha=1}^m (\theta ^\a- \theta^\a(\tau)) 
\Big(1 - \sum_\beta \dot \theta^\b(\tau) d\tau \iota_\b\Big) \prod_{\alpha=1}^m \delta(\psi^\a) \wedge {\mathbb Y}^{(n|0)}_\Lambda  \nonumber \\
&& \hspace{-0.5cm} = \int_{{\cal SM} \times {\mathbb T}} \Big(A_{a}(x, \theta(\tau)) V^a + A_\a(x, \theta(\tau)) \psi^\a\Big) \nonumber \\
&\qquad& \qquad \wedge \, \prod_{\alpha=1}^m (\theta ^\a- \theta^\a(\tau))  \Big(1 - \sum_\beta \dot \theta^\b(\tau) d\tau \iota_\b\Big) \prod_{\alpha=1}^m \delta(\psi^\a) \wedge {\mathbb Y}^{(n|0)}_\Lambda\nonumber 
\end{eqnarray}
where we have used the product $\prod_{\alpha=1}^m (\theta^\alpha - \theta^\alpha(\tau))$ to localize the superfield $\theta$-coordinates on the supercurve.  
Due to the presence of the factor $\prod_\a \delta(\psi^\a)$ the only non-vanishing contributions come from terms in the integrand which do not contain any power of $\psi^\a$, like for instance $A_{a}(x, \theta(\tau)) dx^a$ from the first term, or terms linear in $\psi^\a$ where the action of the contraction $\iota_\a$ has the effect to 
replace $\psi^\a \to \dot\theta^\a d\tau$. Therefore, using the PCO \eqref{BKA2} to localize also the bosonic coordinates on the supercurve $\Lambda$, from eq. \eqref{BKB} we easily find
\begin{eqnarray}
\label{BKC}
\Gamma \! &=& \!  \int_{{\cal SM} \times {\mathbb T}} \! \Big(A_{a}(x, \theta(\tau)) (dx^a \! + \! \theta \gamma^a \dot \theta d\tau) \! + \! A_\a(x, \theta(\tau)) \dot\theta^\a d\tau \Big) \prod_{\alpha=1}^m (\theta ^\a \! - \! \theta^\a(\tau)) \prod_{\alpha=1}^m \delta(\psi^\a) \! \wedge \! {\mathbb Y}^{(n|0)}_\Lambda \nonumber \\
&=& \int_\Lambda  \Big(A_{a}(\tau) \Pi^a(\tau) + A_\a(\tau)  \dot\theta^\a (\tau) \Big) d\tau 
\end{eqnarray}
In the special case of ten dimensional $N=1$ superspace, this expression coincides with (\ref{susyW}), (\ref{susyW2}) and describes the supersymmetric Wilson operator studied in \cite{Ooguri:2000ps, Beisert:2015jxa}. Similarly, in the four dimensional $N=1$ case $\Gamma$ reduces to the well-known superholonomy and gives rise to the super Wilson loop proposed in \cite{Gates:1976rk}\footnote{An alternative construction of abelian supersymmetric Wilson loops has been proposed in \cite{Awada:1995fe}, in terms of superfield strengths rather than superconnections. The two formulations should be related by a super-Stokes theorem in analogy with what happens in the bosonic case (see  eq. \eqref{WLB}).}.

\vskip 12pt
{\bf Properties of the fermionic PCO.} The fermionic PCO defined in \eqref{AKAEgeneral}  satisfies the same properties of the bosonic one, eqs. \eqref{WLE}. Therefore the total operator ${\mathbb Y}^{(n|m)}_\Lambda$ is closed, but not exact and its variations are $d$-exact. 

The last statement is a consequence of a remarkable feature of the fermionic PCO's: Given the non-supersymmetric PCO 
\begin{eqnarray}\label{BKAAA}
{\mathbb Y}^{(0|m)}_0 =   \theta^m \, \d^{(m)}(\psi  )
\end{eqnarray}
corresponding to immersion functions $g^\a(\tau) = \theta^\a$ (i.e. $\theta^\a(\tau) =0$),  then describing an ordinary curve localized at $\theta^\a=0$,  all the fermionic PCO's turn out to be in the same $d$-cohomological class of ${\mathbb Y}^{(0|m)}_0$. 
In order to prove this property we consider a generic ${\mathbb Y}^{(0|m)}_\Lambda$ as given in eq. \eqref{BKA}. Restricting to the the simplest case of a single fermionic dimension ($m=1$), and using $d\theta \delta'(d\theta) = - \delta(d\theta)$,
$d\theta \delta''(d\theta) = - 2 \delta'(d\theta)$, we can write the following chain of identities
\begin{eqnarray}
\label{REMA}
{\mathbb  Y}^{(0|1)}_\Lambda &=& (\theta - \theta(\tau)) \delta\Big(d\theta - \dot \theta(\tau) d\tau\Big) = 
 (\theta - \theta(\tau))\Big(\delta(d\theta) - \dot \theta(\tau) d\tau \delta'(d\theta)\Big) \nonumber \\
&=& \theta \delta(d\theta) - \theta(\tau) \delta(d\theta) - \theta \dot \theta(\tau) d\tau \delta'(d\theta) + \theta(\tau) \dot \theta(\tau) d\tau \delta'(d\theta) \nonumber \\
&=& \theta \delta(d\theta) - d\left[ 
\theta(\tau) \Big( \theta \delta'(d\theta) + \frac12 \dot \theta(\tau) d\tau \theta \delta''(d\theta) \Big) \right] \nonumber \\
&=& {\mathbb  Y}^{(0|1)}_0 + d\hspace{-0.1cm}-\hspace{-0.1cm}{\rm exact ~ term}
\end{eqnarray}
so proving the property in the $m=1$ case. Since the  generalization of the proof to more than one fermionic coordinate is straightforward, we conclude that a generic fermionic PCO is $d$-equivalent to the non-supersymmetric one, independently of the particular defining function $g^\a(x,\theta)$.  It then follows that any pair of PCOs that differ for the choice of the supercontour, i.e. for the choice of the immersion functions, are $d$-equivalent (clearly, if the two contours are linked by a deformation that does not cross singularities). In particular, this implies that the any variation of the PCO induced by a deformation of the path is $d$-exact, as stated above. The same conclusions remain true when we complete the PCO with its bosonic part ${\mathbb Y}^{(n|0)}_\Lambda$. 

Although the addition of $d$-exact terms does not change the cohomological properties of a PCO, it can change its degree of supersymmetry, that is the number of supercharges under which the operator is invariant. We now elaborate on this important point. 

Using the geometrical approach, a supersymmetry transformation generated by a spinor $\e$ acts on the PCO as the Lie derivative \eqref{Lieder}. Exploiting its $d$-closure property we can write 
\begin{eqnarray}
\delta_\epsilon {\mathbb Y}^{(n|m)}_\Lambda =  d \iota_\epsilon  {\mathbb Y}^{(n|m)}_\Lambda 
\end{eqnarray}
The ${\mathbb Y}^{(0|m)}_0$ operator introduced above breaks supersymmetry completely, $\delta_\epsilon {\mathbb Y}^{(0|m)}_0 \neq 0$. In fact,  its defining constraints $\theta^\a = 0$ are trivially not invariant under supersymmetry transformations, $\d \theta^\a = \e^\a$. However, we can perform the shift (we include also the bosonic part)
\begin{eqnarray}\label{shift0}
{\mathbb Y}^{(n|m)}_0  \to {\mathbb Y}^{(n|m)}_{\Lambda} = {\mathbb Y}^{(n|m)}_0  +  d\Sigma^{(n-1|m)}
\end{eqnarray}
and determine $\Sigma^{(n-1|m)}$ in such a way that $\d_\e {\mathbb Y}^{(n|m)}_{ \Lambda} =0$. This condition is equivalent to requiring 
\begin{eqnarray}
\label{AKAPA}
\delta_\epsilon \Sigma^{(n-1|m)} = - \iota_\epsilon  {\mathbb Y}^{(n|m)}_0
\end{eqnarray}
If this equation is true for arbitrary $\e^\a$, then the shifted PCO is manifestly invariant under all the supersymmetry charges. In general the Killing spinors $\e$ are a function of $\tau$ and supersymmetry is realized {\em locally} on the contour.  

Therefore, by simply adding a $d$-exact term we can move from a PCO localizing on a non-supersymmetric contour to a PCO localizing on a supersymmetric one. Between these two extreme cases we may have a pletora of intermediate situations where eq. \eqref{AKAPA} holds only for a subset of Killing spinor $\e^\a$ components, so defining PCO localizing on partially supersymmetric supercontours. An easy way to convince about this fact is to consider for instance the following PCO in ten dimensions
\begin{eqnarray}
\label{newPiA}
\mathbb{Y}^{(0|16)} = \epsilon_{\a_1 \dots \alpha_{16}} \theta^{\alpha_1} \dots \theta^{\alpha_{15}} (V_a \gamma^a \iota)^{\alpha_{16}} \delta^{16}(d\theta)
\end{eqnarray}
obtained from the non-supersymmetric $\mathbb{Y}^{(0|16)}_0 = \epsilon_{\a_1 \dots \alpha_{16}} \theta^{\alpha_1} \dots \theta^{\alpha_{16}} \delta^{16}(d\theta)$
by replacing $\theta^{\a_{16}}$ with the supersymmetric expression $(V_a \gamma^a \iota)^{\alpha_{16}} $. Writing $V^a$ explicitly as in \eqref{supervielbeins}, after little algebra one can show that this PCO is $d$-equivalent to the non-supersymmetric one
\begin{eqnarray}
\label{newPiB}
\mathbb{Y}^{(0|16)} =   \mathbb{Y}^{(0|16)}_0  - 
d \left[  \epsilon_{\a_1 \dots \alpha_{16}} \theta^{\alpha_1} \dots \theta^{\alpha_{15}} 
x_a (\gamma^a \iota)^{\alpha_{16}} \delta^{16}(d\theta)
\right] 
\end{eqnarray}
and is invariant under a supersymmetry transformation generated by the Killing spinor $\e = (1, 0, \dots , 0)$, that is it preserves only one supercharge.  More generally, if in $\mathbb{Y}^{(0|16)}_0$ we replace $\theta^{\a_1} \dots \theta^{\a_p}$ with $p$ factors $(V_a \gamma^a \iota)^{\alpha_{i}} $ we obtain a well-defined fermionic PCO which preserves $p$ supercharges. We note that this procedure can be applied as long as $p \leq 10$. Beyond that limit, we would end up with an exceeding number of $V$ forms that would trivialize the expression. In particular, this construction cannot be used to generate a fully supersymmetric PCO.

In the fully supersymmetric case, we claim that the solution to \eqref{AKAPA} is given by the following expression 
\begin{eqnarray} \label{AKAPAA}
 {\mathbb{Y}}^{(n|m)}_{  \Lambda} &=& \delta^n \left( x^a - x^a (\tau) - \left( V_b - \Pi_b d \tau \right) \left( \theta - \theta (\tau) \right) \gamma^{ab} \iota \right) \left( V - \Pi d \tau \right)^n   \\
 && \qquad  \qquad \wedge \left( \theta - \theta \left( \tau \right) - \left( dx^a - \dot{x}^a d \tau \right) \gamma_a \iota \right)^m \delta^m \left( d \theta - \dot{\theta} d \tau \right) \nonumber \\
	&=&  e^{ - \mathcal{L}_{\partial_\tau} } \left[ \delta^n \left( x^a - V_b \theta \gamma^{ab} \iota \right) V ^n \left( \theta - dx^a \gamma_a \iota \right)^m \delta^m \left( d \theta \right) \right] \equiv  e^{ - \mathcal{L}_{\partial_\tau} } \mathbb{Y}' \nonumber
\end{eqnarray}
where we have introduced the Lie derivative along the vector field $\partial_\tau$, the tangent vector along the curve. 

In order to support this statement we prove that \eqref{AKAPAA} is $d$-closed and invariant under supersymmetry transformations. To this end, it is convenient to remind the following identities
\begin{eqnarray}
	\label{AKAPAC} \left[ d , \mathcal{L}_{\partial_\tau} \right] = 0 \qquad , \qquad \left[ \mathcal{L}_\epsilon , \mathcal{L}_{\partial_\tau} \right] = \mathcal{L}_{\left[ \epsilon , \partial_\tau \right]} = \mathcal{L}_{-\dot{\epsilon}^\alpha Q_\alpha} = \mathcal{L}_{-\dot{\epsilon}}  
	\end{eqnarray}
which easily imply 	
\begin{eqnarray}
	\label{AKAPAC2}
	 d \left( e^{ - \mathcal{L}_{\partial_\tau}} \mathbb{Y}' \right) =  e^{ - \mathcal{L}_{\partial_\tau} } d  \mathbb{Y}' 
 , \qquad  \mathcal{L}_\epsilon \exp \left( - \mathcal{L}_{\partial_\tau} \right) \mathbb{Y}' = \exp \left( - \mathcal{L}_{\partial_\tau} \right) \mathcal{L}_\epsilon  \mathbb{Y}' + \mathcal{L}_{\tilde{\epsilon}} \mathbb{Y}' 
\end{eqnarray}
Here we have introduced the super-vector field $\tilde{\epsilon} = ( 1 - \exp ( - \partial_\tau ) ) \epsilon^\alpha Q_\alpha$. We note that this super-vector is vanishing in the case of supersymmetry globally defined on the supercontour.
From eqs. \eqref{AKAPAC2} it then follows  that it is sufficient to study the closure and the supersymmetry invariance of $\mathbb{Y}'$. For sake of clarity, we do the calculation in the simplest case of $n=m=1$, being the generalisation lengthy but straightforward.  For the $d$-closure we have
\begin{eqnarray}
	\nonumber d \mathbb{Y}'  &=& d \left[ \delta ( x - dx \theta \iota ) V ( \theta - dx \iota  ) \delta ( d \theta  ) \right] = \delta'  ( x - dx \theta \iota )  ( dx + dx d \theta \iota ) V  ( \theta - dx \iota  ) \delta  ( d \theta  ) + \\ &&+ \delta ( x - dx \theta \iota ) ( d \theta )^2  ( \theta - dx \iota) \delta ( d \theta ) + \delta  ( x - dx \theta \iota) V d \theta \delta ( d \theta ) = 0 
\end{eqnarray}
whereas for the supersymmetry variation we obtain
\begin{eqnarray}
	\nonumber \delta_\epsilon \mathbb{Y}' &=& \delta_\epsilon \left[ \delta \left( x - dx \theta \iota \right) V \left( \theta - dx \iota \right) \delta \left( d \theta \right) \right] = \delta' \left( x - dx \theta \iota \right) \epsilon \left( \theta + V \iota \right) V \left( \theta - dx \iota \right) \delta \left( d \theta \right) + \\ && + \delta \left( x - dx \theta \iota \right) V \left( \epsilon + \epsilon d \theta \iota \right) \delta \left( d \theta \right) = 0
\end{eqnarray}
and the same for $\d_{\tilde \e} \mathbb{Y}' $. The results have been obtained by using nilpotence properties like $\theta^2 = 0 = dx \wedge dx$ and the usual distributional properties recalled in section 2.
Now, inserting back in \eqref{AKAPAC2} we conclude that ${\mathbb{Y}}^{(n|m)}_{  \Lambda}$ is indeed closed and fully supersymmetric. 

To close this section it is important to observe that if two PCO's correspond to two different supercontours, and therefore differ by a $d$-exact term, they give rise in general to two different Wilson operators. In fact, if we start from \eqref{AKAF} and perform the shift $ {\mathbb Y}^{(n-1|m)} \rightarrow  {\mathbb Y}^{(n-1|m)} + d \Sigma^{(n-2|m)}$  
the $\Gamma$ integral undergoes the  following non-trivial change
\begin{eqnarray}
\label{AKAL}
\Gamma \to \Gamma' = \int_{\cal SM} A^{(1|0)} \wedge \Big( {\mathbb Y}^{(n-1|m)} + d  \Sigma^{(n-2|m)} \Big) 
=   \Gamma + \int_{\cal SM}  F^{(2|0)} \wedge \Sigma^{(n-2|m)} 
\end{eqnarray}
 where $F^{(2|0)}= dA^{(1|0)}$ is the field-strength which is in general non-vanishing on ${\cal SM}$. 
Therefore, by tuning the $d$-exact term we can flow from one operator to another one. In particular, since different choices of PCO's may correspond to different degrees of supersymmetry preserved by the corresponding supercontours, the $d$-cohomological equivalence can be used to vary the number of supercharges preserved by the Wilson loop. This will be discussed in detail in section \ref{sect:susy}, whereas in the next subsection we give a first example of this mechanism at work.

\subsection{The Wilson-Maldacena Operator in $N=4$ SYM Theory}\label{WM}

In this section we provide an explicit example  of the {\it $d$-varying supersymmetry} mechanism described above  by studying the remarkable case of the Wilson-Maldacena loop in four dimensional $N=4$ SYM theory  \cite{Maldacena:1998im, Rey:1998ik}. 

We consider the four dimensional $N=4$ SYM theory formulated in the $(4|16)$-supermanifold. An ordinary Wilson loop along a curve $\lambda$ parametrized by $\tau \to x^a(\tau)$,  is defined as in eq. \eqref{BKB} by taking the non-supersymmetric PCO
\begin{eqnarray}
{\mathbb  Y}_0^{(4|16)} = \prod_{a=1}^4 \delta\Big(x^a - x^a(\tau) \Big) \bigwedge_{a=1}^4 (dx^a - \dot{x}^a d\tau) \,  \prod_{\alpha=1}^{16} \theta^\a \delta(\psi^\a)
\end{eqnarray} 
As already observed, it never preserves any supercharge, no matter is the choice of the contour. Instead, let us consider the $d$-equivalent PCO
 \begin{eqnarray}\label{shift}
{\mathbb  Y}^{(4|16)}  = {\mathbb  Y}_0^{(4|16)} + d \, \Sigma^{(3|16)}
\end{eqnarray}
with 
\begin{eqnarray}
\label{pippoP}
\Sigma^{(3|16)} &=& d\tau  \prod_{\rho=1}^{16}\Big(\theta^\rho - \theta^\rho(\tau)\Big) 
\prod_{a=1}^4 \delta\Big(x^a - x^a (\tau)\Big)  
\epsilon_{a_1 \dots a_4} V^{a_1} \dots V^{a_4} \nonumber \\
&\times& \Big( N^{AB} \epsilon^{\a\b} 
\iota_{\alpha A} \iota_{\beta B} + \bar N_{AB}  \epsilon^{\dot\a\dot\b} 
\iota_{\dot\alpha}^A \iota_{\dot\beta}^B
\Big)\delta^{16} (\psi) 
\end{eqnarray}
Here $\iota_\alpha$ is the contraction respect to fermionic vector field $\partial_\a$, and $ N_{AB}$ is a real vector of the $SU(4)$ R-symmetry group satisfying $\bar N_{AB} = \epsilon_{ABCD} N^{CD}$.

Plugging the shifted PCO \eqref{shift} into the general expression for $\G$  we obtain a shifted holonomy of the form \eqref{AKAL}.
If we now replace $F^{(2|0)}$ with its explicit expression \eqref{decoC} valid for the $N=4$ case, thanks to its non-trivial dependence on the scalar fields, we obtain
\begin{eqnarray}
\label{pippoQ}
\Gamma =  \int_\lambda \left( A_a \dot{x}^a +N^{AB} \bar \phi_{AB} + \bar N_{AB} \phi^{AB} \right) d\tau
\end{eqnarray}
This expression coincides with the integral of the Wilson-Maldacena generalised connection that includes non-trivial couplings to the six scalars $\phi^{[AB]}$.  As is well-known, under a suitable choice of the $\l$ contour  and the internal couplings $N_{AB}$ this operator is partially supersymmetric \cite{Drukker:1999zq}. 
Therefore, this example proves that $d$-exact terms can be used to enhance the degree of supersymmetry of a Wilson operator. 

More generally, if we start from the super-Wilson loop \eqref{BKB} corresponding to a generic PCO \eqref{BKA3} and perform the shift  ${\mathbb  Y}^{(4|16)}_\L  \to  {\mathbb  Y}_\L^{(4|16)} + d \, \Sigma^{(3|16)}$, with a similar procedure we find the supersymmetric version of the Wilson-Maldacena operator
\begin{eqnarray}
\label{pippoQ2}
\Gamma =  \int_\L \left( A_a \Pi^a + A_\a \dot{\theta}^\a + N^{AB} \bar \Phi_{AB} + \bar N_{AB} \Phi^{AB} \right) d\tau
\end{eqnarray}
which has been proposed in \cite{Beisert:2015jxa}. 

This construction holds for any gauge theory with extended supersymmetry $N\geq 2$. In fact, in all these cases the superfield strength $F^{(2|0)}$ contains terms of the form $F_{\a I \b J} \psi^{\a I } \psi^{\b J}$, with $F_{\a I \b J}$ being proportional to the scalar 
fields of the gauge multiplet \cite{cube}.  Therefore, as in the Wilson-Maldacena example, a careful choice of $\Sigma^{(n-1|m)}$ leads to an operator which contains non-trivial couplings to the scalar sector.

\sect{Variations and Symmetries} \label{sect:variations}

In this section we study how invariances of a super-Wilson loop can be studied in the language of supermanifolds.  As representatives we will consider operators in $N=1$ SYM in ten dimensions and $N=4$ SYM in four dimensions.  We begin by checking invariance under a reparametrization of the path, and then move to the study of invariance under superdiffeomorphisms, supersymmetry and kappa symmetry. 

\subsection{Reparametrisation Invariance of the PCO}

We start by briefly studying the reparametrisation invariance of the PCO in \eqref{BKA3}. To this end, it is convenient to rewrite it in the following form
\begin{equation}\label{RIB}
	{\mathbb Y}^{(n|m)}_\Lambda = \left( \iota_\tau + \dot{\theta}^\alpha \iota_\alpha + \Pi^a \iota_a \right) \text{Vol} 
\end{equation}
where we have introduced the volume form
\begin{equation}\label{RIA}
\text{Vol} =  \delta^{(n)} \left( x-x(\tau) \right)  V^{n}  d \tau \times \left( \theta - \theta(\tau) \right)^{m} \delta^{(m)} \left( \psi \right) 
\end{equation}
Now, under a given reparametrisation $\tau \mapsto \sigma (\tau)$, the PCO variation, expressed as usual by a Lie derivative, reads
\begin{equation}\label{RIC}
	\delta_\sigma {\mathbb Y}^{(n|m)}_\Lambda = d \iota_\sigma {\mathbb Y}^{(n|m)}_\Lambda = d \left[ \sigma \left( \iota_\tau + \dot{\theta}^\alpha \iota_\alpha + \Pi^a \iota_a \right) {\mathbb Y}_\Lambda \right] = d \left[ \sigma \left( \iota_\tau + \dot{\theta}^\alpha \iota_\alpha + \Pi^a \iota_a \right)^2 \text{Vol} \right] = 0 
\end{equation}
since the object inside the round brackets is odd. This proves the independence of the $\Gamma$ integral from the contour parametrization.

\subsection{Variation under Superdiffeomorphisms}\label{superdiff}

Given a super-Wilson loop ${\cal W} = e^{\Gamma}$ with $\Gamma$ written as in eq. \eqref{AKAF}, we study its behavior under an infinitesimal superdiffeomorphism generated by a vector field $X$. This is equivalent to studying how the $\G$ exponent transforms. Since we have written $\G$ as a top form integrated on the entire supermanifold and a generic superdiffeomorphism is nothing but a change of coordinates in the supermanifold, we can immediately conclude that by construction $\G$, and then ${\cal W}$, are manifestly invariant under superdiffeomorphisms. Explicitly, taking into account that for an infinitesimal trasformation the PCO changes by a $d$-exact term, $\delta_X  {\mathbb Y}^{(n|m)}_\Lambda  = d \iota_X  {\mathbb Y}^{(n|m)}_\Lambda$, we can write 
\begin{eqnarray}
\label{varA}
 \d_X \Gamma  = \int_{\mathcal {SM} \times {\mathbb T}} \left(  \iota_X F^{(2|0)}  \wedge {\mathbb Y}^{(n|m)}_\Lambda + A^{(1|0)} \wedge d \iota_X  {\mathbb Y}^{(n|m)}_\Lambda\right) \equiv 0
\end{eqnarray}
If in the second term we integrate by parts and assume that there are no boundary terms, this identity can be equivalently written as
\begin{eqnarray}
\label{varC}
\iota_X F^{(2|0)}  \wedge  {\mathbb Y}^{(n|m)}_\Lambda + F^{(2|0)}   \wedge \iota_X  {\mathbb Y}^{(n|m)}_\Lambda = d \Omega^{(n|m)}
\end{eqnarray}
for any arbitrary $\Omega^{(n|m)}$ form. 

Identity \eqref{varA} is equivalent to state that in superspace the variation in form of the superconnection induced by the $X$-tranformation is compensated by the variation of the supercontour $\Lambda$ encoded in the PCO. In other words, we can write 
\begin{eqnarray}\label{varB}
(\d_X\G)(\L) = - \G(\d_X \L)
\end{eqnarray}
where $\d_X$ on the l.h.s. is the $X$-variation done by keeping the supercontour fixed\footnote{Here we use the same symbol $\d_X$ to indicate both the variation in form of the fields  and the variation of the coordinates of the supermanifold.}. When uplifted at the level of the super-Wilson loop, taking into account that a PCO identifies a supercontour uniquely, this implies that $(\d_X {\cal W})(\L) = - {\cal W}(\d _X \L)$. Therefore, the variation of the Wilson operator follows from the $X$-transformation of the supercontour. In particular, a given $X$-diffeomorphism is a symmetry for ${\cal W}$ if $(\d_X \G)(\L) = 0$, but from identity \eqref{varB} this is true if and only if $\d_X \L = 0$. Therefore, the set of ${\cal W}$ invariances coincides with the set of $\L$ symmetries. We note that the same reasoning can be applied to bosonic loops defined in ordinary manifolds: $(\d_X W)(\l)=0$ if and only if $\d_X \l = 0$.

\subsection{Supersymmetry Invariance}\label{sect:susy}

A supersymmetry transformation is a particular superdiffeomorphism generated by the vector $X \equiv \e = \e^\a Q_\a$, where $Q_\a$ are the supersymmetry charges. Therefore, the behavior of a Wilson loop under supersymmetry transformations can be easily infered from the discussion  in the previous section. In particular, $\G$ is manifestly supersymmetric by construction, and from \eqref{varB} we can write $(\d_\e \G)(\L) = -\G({\d_\e \L})$. This means that its variation is entirely due to the variation of the supercontour. 
 This property has been already discussed in \cite{Muller:2013rta, Beisert:2015jxa}. What is interesting to stress here is that in the present formalism, being the $\Gamma$'s integrand factorized into the product of a contour-independent superfield and a PCO that encloses the whole dependence on the contour, this pattern arises straightforwardly.

A Wilson loop preserves a given amount of supersymmetry (it is BPS) when for a particular generator $\e$ it satisfies $(\d_\e {\cal W})(\L) =0$, or equivalently $(\d_\e \G)(\L) =0$. But, from the previous reasoning this can be traded for the condition
$\G({\d_\e \L})=0$. Therefore, counting the number of supersymmetries preserved by ${\cal W}$ gets translated into counting the number of supersymmetries preserved by the corresponding supercontour. More precisely, from \eqref{varC} we read
\begin{eqnarray}\label{invariance}
(\d_\e \G)(\L) = 0    \quad \Longleftrightarrow  \quad F^{(2|0)} \wedge \iota_\e {\mathbb Y}_\L^{(n|m)} = 0
\end{eqnarray}
up to $d$-exact terms that we neglect.  
 
 As discussed in section \ref{sect:susyWL}, we can exploit the $d$-equivalence of super-PCO's to vary their degree of supersymmetry. Precisely, given a particular supersymmetry transformation generated by an assigned $\e$ we can always construct an $\e$-preserving PCO from an $\e$-breaking operator by performing the shift \eqref{shift0}, with $\S^{(n-1|m)}$ satisfying condition \eqref{AKAPA}. Therefore, choosing a specific representative within the $d$-class corresponds to fixing the amount of supersymmetry preserved by the corresponding  Wilson loop. Enhancing or de-enhancing supersymmetry can then be done by adding $d$-exact terms. This result may have important implications in the study of renormalization group flows between Wilson operators preserving different amount of supersymmetry \cite{Polchinski:2011im,Beccaria:2017rbe,Beccaria:2018ocq}. 

Equation \eqref{invariance} is the Killing spinor equation selecting the supersymmetry invariances of an assigned Wilson operator. We study it in details, in the ten dimensional case.

First of all, if we express the PCO as in eq. \eqref{BKA3} and take into account identities \eqref{AKAK}, the $\iota_\epsilon$-contraction on ${\mathbb Y}^{(10|16)}_\Lambda$ gives rise to the following two terms
\begin{eqnarray}\label{twoterms}
	\nonumber \hspace{-1cm} \iota_\epsilon {\mathbb Y}^{(10|16)}_\Lambda &=& \d^{(10)} (x - x(\tau)) \, 2 \epsilon \gamma^a \theta \iota_a ( V - \Pi(\tau) d\tau)^{10} \, \wedge \, (\theta - \theta(\tau) )^{16} \, \d^{(16)}(\psi - \dot{\theta} (\tau) d\tau ) \\
	\label{CAA} &+& \d^{(10)} (x - x(\tau)) \, ( V - \Pi(\tau) d\tau)^{10} \, \wedge \, (\theta - \theta(\tau) )^{16} \, \epsilon^\alpha \iota_\alpha \d^{(16)}(\psi - \dot{\theta} (\tau) d\tau )
\end{eqnarray}
Now, according to \eqref{invariance}, this expression has to be multiplied by $F^{(2|0)}$. Using the rheonomic parametrization \eqref{SYMB}, it is easy to see that from the first term in \eqref{twoterms} we obtain a non-trivial contribution both from $F_{ab} V^a V^b$ and $(\psi \g_a W )V^a$, whereas from the second term we obtain only one contribution from $(\psi \g_a W )V^a$, being the $VV$ term trivially zero. 
Summing all the contributions and factorizing out the volume form \eqref{RIA}, we finally obtain that the Killing spinor equation reads
\begin{equation}\label{CAE}
\left( 2 \epsilon \gamma^a \theta \Pi^b F_{ab} - 2 \epsilon \gamma^a \theta  \, W \gamma_a \dot{\theta} + \epsilon \gamma_a W \Pi^a \right) \hspace{-0.1cm} \Big|_{\L}= 0
\end{equation}
where all the quantities are evaluated on the supercontour. 

When we deal with a supersymmetry preserving PCO,  identity \eqref{varC} implies that the following equation
\begin{eqnarray} \label{killing2}
 \iota_\epsilon F^{(2|0)} \wedge {\mathbb Y}^{(10|16)}_{\Lambda} = 0 
\end{eqnarray}
has to be automatically satisfied, up to $d$-terms. There are two possibilities for which this is true.
Exploiting the $d$-closure of the PCO, the first possibility is that $\iota_\e F^{(2|0)}  = d \Upsilon^{(0|0)}$
on the entire supermanifold, or the even stronger condition $\iota_\e F^{(2|0)} =0$.
These conditions imply a constraint on the gauge field itself and are rarely satisfied\footnote{We note that this is generically what happens 
for a supersymmetric invariant action, $\iota_\epsilon d {\cal L}^{(n|0)} =0$. If the action is $d$-closed (which is possible when auxiliary fields are present), then we have a manifest supersymmetric action. In other cases, the absence of auxiliary fields implies that the action satisfies the weaker condition.}. 
The second possibility is that
\begin{eqnarray}
\label{conB}
\iota_\e F^{(2|0)} \in {\rm ker} \, {\mathbb Y}^{(10|16)}_{ \Lambda} 
\end{eqnarray}
up to $d$-terms, which means that $\iota_\epsilon F^{(2|0)}$ is vanishing or it is a total derivative {\em on the supercontour only}. 
Using the explicit expression \eqref{SYMB} for the superfield strength it is easy to check that this condition leads exactly to the Killing spinor equation 
\eqref{CAE}. This is a consistency check of the manifest supersymmetry invariance in superspace.
 
In general, for arbitrary values of the field strengths, equation \eqref{CAE} can be solved locally on the contour, leading to a \emph{local supersymmetry} generated by a Killing spinor $\e(\tau)$. Remarkably, in the case of a Wilson loop defined on an ordinary bosonic path ($\theta(\tau)=0$ on the supercontour) it leads to the well-known condition
\begin{eqnarray}\label{conD}
\epsilon(\tau)  \gamma_a \dot{x}^a(\tau) = 0
\end{eqnarray}
When reduced to four dimensions, solutions to this equation for $\e$ constant  lead to  Zarembo-like BPS operators in $N=4$ SYM \cite{zarembo}. Instead, in the case of ten dimensional light-like paths, eq. \eqref{conD} has a non-trivial kernel, since it automatically squares to 0. Reduced to four dimensions it defines 1/2-BPS operators in $N=4$ SYM if the extra coordinates are identified with the internal couplings to the scalars \cite{Maldacena:1998im,Drukker:1999zq}. In this case a systematic classification of solutions to \eqref{conD} has been given in \cite{Dymarsky:2009si}, which involves ten dimensional pure spinors. 
We note that the light-like nature of the contour in ten dimensions is related to kappa-symmetry, as we are going to analyse in the next section.

\subsection{Kappa Symmetry}

The superconnection $\G$ that defines a Wilson loop can be interpreted as the action of a non-dynamical superparticle moving in an electromagnetic field. Since the superparticle in ten dimensions exhibits kappa-symmetry invariance \cite{Siegel:1983hh}, it is sensible to study how the ten dimensional $\G$ behaves under this symmetry. This has been extensively discussed in \cite{Ooguri:2000ps, Beisert:2015jxa, Beisert:2015uda}. Here we reformulate the problem in the language of superdifferential forms.  In particular, we will confirm the result that kappa-symmetry invariance in ten dimensions is strictly related to BPS properties of the super-Wilson operator in $N=4$ SYM theory. 

A kappa-symmetry transformation is generated by a vector $\widetilde \kappa \equiv \kappa^\a D_\a$, with the kappa-symmetry parameter expressed in terms of geometric data as
\begin{equation}\label{KSF}
	\kappa^\alpha = ( \gamma^a )^{\alpha \beta} \mathcal{L}_a K_\beta 
\end{equation}
Here $K_\b$ is a $0$-form carrying a spinorial index and  $\mathcal{L}_a$ is the infinitesimal translation operator. As is well-known, only half of the $\kappa^\a$ components are independent. This can be easily understood by proving that the operator $\left( \gamma^a \right)^{\alpha \beta} \mathcal{L}_a$ has a non-trivial kernel, thus allowing to fix half of the fermionic components. An alternative proof, as well as kappa-symmetry transformations of the coordinates, of the basic one-forms and of generic superfields are reviewed in appendix \ref{app1}.

\subsubsection{Kappa-symmetry for the super-Wilson loop in 10D}

We investigate the action of kappa-symmetry on the Wilson operator ${\cal W} = e^\G$, with $\G$ given in \eqref{BKA3}. 
Since kappa-symmetry transformations fall into the class of superdiffeomorphisms discussed in section \ref{superdiff} the Wilson loop is manifestly invariant under kappa-symmetry by construction. In particular, it has to satisfy identity \eqref{varA} with $X=\widetilde{\kappa}$, which once again tells us that the Wilson loop variation is entirely due to the variation of its supercontour, i.e. $(\d_{\widetilde \kappa} {\cal W}) (\L) = - {\cal W}(\d_{\widetilde \kappa} \L)$.

We want to study the WL behavior $(\d_{\widetilde \kappa} {\cal W} )(\L)$ at fixed $\L$ and see under which conditions this variation, or equivalently $(\d_{\widetilde \kappa} \G) (\L)$, vanishes. As just said, and in analogy to what we have done for supersymmetry invariance, this is traded by the following condition
\begin{eqnarray}\label{k-invariance}
 \G( \d_{\widetilde \kappa}\L) = 0    \quad \Longleftrightarrow  \quad F^{(2|0)} \wedge \iota_{\widetilde \kappa} {\mathbb Y}_\L^{(n|m)} = 0
\end{eqnarray}
Specializing to ten dimensions and using identities \eqref{kappaAB} we have
\begin{equation}\label{KSP}
 \iota_{\widetilde \kappa} {\mathbb Y}_\L^{(10|16)}   =  \left( \theta - \theta(\tau) \right)^{16} \delta^{10} ( x-x(\tau)) \left( V - \Pi d \tau \right)^{10}  \kappa^\alpha \iota_\alpha \delta^{16} ( \psi - \dot{\theta} d \tau ) 
\end{equation}
The wedge product with $F^{(2|0)}$ in \eqref{SYMB} eventually gives
\begin{eqnarray}\label{KSR}
F^{(2|0)} \wedge  \iota_{\widetilde \kappa} {\mathbb Y}_\L^{(10|16)}   = - 10 \, \Pi^a (W \gamma_{a}  \kappa) \times \text{Vol}  
\end{eqnarray}
with the volume form given in \eqref{RIA}. Integrating on ${\cal SM} \times {\mathbb T}$ we eventually obtain that invariance under kappa-symmetry transformations is ensured by the condition 
\begin{equation}\label{KSQ}
\Pi^a (W \gamma_{a}  \kappa)\Big|_{\L} = 0 
\end{equation}
Substituting $\kappa$ with expression \eqref{KSF} in momentum representation and localised on the supercontour we end up with $\delta_{\widetilde \kappa} \G \propto  \Pi^2$. Therefore, the Wilson loop invariance under kappa-symmetry is ensured by the light-like condition, $\Pi^2 (\tau) = 0$ at each point of the contour. In the AdS/CFT framework, the worldline kappa-symmetry invariance of the Wilson loop corresponds to the kappa-symmetry invariance of the dual string worldsheet \cite{Ooguri:2000ps}. 

We note that the fact that we are in ten dimensions has not played any special role in the derivation of this result. Therefore, the same procedure can be applied to super-Wilson loops in 4D without the Wilson-Maldacena terms. Also in that case we find that kappa-symmetry is ensured by the light-like condition on the supercovariant momentum \cite{Beisert:2012gb}. 

\subsubsection{Kappa-symmetry for the Wilson-Maldacena loop in 4D }

We now study the kappa-symmetry variation of the four dimensional Wilson-Maldacena connection given in eq. \eqref{pippoQ}. As discussed in section \ref{WM} this connection can be written in terms of an integrable superform associated to PCO (\ref{shift}), (\ref{pippoP}), which differs from the PCO localizing the path at $\theta^\a = 0$ by a $d$-exact term. Explicitly it is given by
\begin{eqnarray}\label{KS2B2}
\hspace{-0.7cm} {\mathbb Y}^{(4|16)}_\L = {\mathbb Y}^{(4|16)}_0 + d \S^{(3|16)}  
= {\mathbb Y}^{(4|16)}_0 + 
\left( N^{AB} \epsilon^{\a\b} D_{\alpha A}  \iota_{\beta B} +
\bar N_{AB} \epsilon^{\dot\a\dot\b} \bar D_{\dot\alpha}^A  \iota_{\dot\beta}^B \right) \times \text{Vol}  
\end{eqnarray}
where $D_{\a A}, \bar D_{\dot\a}^A$, $A=1, \dots, 4$ are the covariant spinorial derivatives in the non-chiral $N=4$ superspace. 

As discussed above, the kappa-symmetry invariance of the corresponding Wilson loop is ensured when the form
\begin{equation}\label{KS2A}
F^{(2|0)} \wedge \iota_{\widetilde \kappa} {\mathbb Y}^{(4|16)}_\L = F^{(2|0)} \wedge \iota_{\widetilde \kappa} {\mathbb Y}^{(4|16)}_0 + 
F^{(2|0)} \wedge \iota_{\widetilde \kappa} d \S^{(3|16)}  
\end{equation}
is integrated to zero. 

Being the first term in \eqref{KS2A} similar to the ten dimensional expression studied in the previous section, its variation can be easily figured out by reducing the previous result \eqref{KSQ} to four dimensions. We obtain
\begin{equation}\label{KS2HA}
	F^{(2|0)} \wedge \iota_{\widetilde \kappa} {\mathbb Y}^{(4|16)}_0 = 4  \,  
	(W^{\a A} \bar\kappa^{\dot\a}_A + \bar W^{\dot\a}_A \kappa^{\a A}) 
\Pi_{\a\dot\a} \times \text{Vol} 
\end{equation}

The second term in \eqref{KS2A} is new and requires a separated  analysis. First of all, neglecting $d$-exact terms, from \eqref{KS2B2} we obtain
\begin{eqnarray}\label{KS2F}
\iota_{\widetilde \kappa} d \S^{(3|16)} 
= 2 \Big[ N^{AB} \epsilon^{\a\b} D_{\alpha A}  \iota_{\beta B} +
\bar N_{AB} \epsilon^{\dot\a\dot\b} \bar D_{\dot\alpha}^A  \iota_{\dot\beta}^B \Big]
\Big(\kappa^{\gamma C} \iota_{\gamma C} + 
 \kappa^{\dot\gamma}_C \iota_{\dot\gamma}^C\Big)
 \times \text{Vol} \nonumber 
\end{eqnarray}
Now taking the wedge product with the superfield strength given in \eqref{decoC}, it is easy to realize that only the last two terms  there contribute and we are left with 
\begin{equation}\label{KS2G}
	F^{(2|0)} \wedge  \iota_{\widetilde \kappa} d \S^{(3|16)} = - 4 \, 
	\Big(
	W^{\alpha A} \kappa^{\beta B}\epsilon_{\a\b} \bar N_{AB} + 
	\bar W^{\dot \alpha}_A \bar\kappa^{\dot\beta}_B \,
	\epsilon_{\dot\a\dot\b} N^{AB} \Big)	
	 \times \text{Vol}
\end{equation}

We now have to sum the two expressions \eqref{KS2HA} and \eqref{KS2G}, and choose a particular parametrization for the four-dimensional spinors in terms of independent components. The most general expression with the correct index structure is
\begin{equation}
\kappa^{\alpha A}  = \Pi^{\a\dot\a} \bar K_{\dot\a}^A + N^{AB} K^\alpha_B \qquad , \qquad \bar\kappa^{\dot\alpha}_A  = \Pi^{\a\dot\a} K_{\a A} + \bar N_{AB} \bar K^{\dot\alpha B}
\end{equation} 
Inserting in the previous equations it is easy to see that mixed $\Pi$-$N$ and $\Pi$-${\bar N}$ contributions cancel, whereas from \eqref{KS2HA} we obtain a term proportional to the four-dimensional $\Pi^2 \equiv \Pi^{\a\dot\a} \Pi_{\a\dot\a}$ and from \eqref{KS2G} an expression proportional to $N^{AB} \bar N_{AB}$. The total variation $\delta_{\widetilde \l} \G$ turns out to be proportional to $(\Pi^2 + N^{AB} \bar N_{AB})$. Therefore, invariance under kappa-symmetry requires $\Pi^2 = - N^{AB} \bar N_{AB}$. This is the well-known condition that in four dimensional $N=4$ SYM  theory leads to BPS Wilson loops \cite{Drukker:1999zq}.

Again, this formalism allows for an easy extension to the general case \eqref{pippoQ2}.

\sect{Generalization to non-abelian gauge groups}\label{nonabelian}

The construction of super-Wilson loops in terms of integral forms can be strightforwadly generalized to the case of a non-abelian gauge theory. In fact, it is sufficient to recall that in the non-abelian case the ordinary definition of a gauge invariant Wilson operator reads
\begin{eqnarray}
\label{NONA}
W = {\rm Tr}_{\cal R} P e^{ \Gamma}\,, ~~~~~~
\Gamma = \oint_\lambda A_*^{(1)}  
\end{eqnarray}
where $\l$ is a closed path, ${\rm Tr}_{\cal R}$ is the trace in representation ${\cal R}$ and the exponential has been generalized to a path ordered 
exponential\footnote{We use the convention 
$P \left( \int_\lambda A_*^{(1)} \right)^n = n! \int_{t_i}^{t_f} dt_1 \int_{t_i}^{t_1} dt_2 \dots \int_{t_i}^{t_{n-1}} dt_n \, A_*^{(1)}(t_1)  A_*^{(1)}(t_2) \dots  A_*^{(1)}(t_n)$.}. 
Therefore, in the present set-up it is sufficient to use definition \eqref{NONA}, but write $\G$ as in \eqref{WLF} for the bosonic operator and \eqref{AKAF} for the supersymmetric one. 

What is interesting to investigate is how in this geometric set-up the invariances of the (super)-Wilson loop discussed in sects. \ref{sect:abelian} and \ref{sect:variations} generalize to the case of a non-abelian (super)connection. As prototypical examples, we are going to study gauge invariance of the bosonic Wilson loop and the conditions for supersymmetry invariance of the super-Wilson operator in ten dimensions.

\subsection{Gauge Invariance}\label{sect:gaugeinvariance}

For a gauge theory associated to a non-abelian group ${\cal G}$, we consider the bosonic $W$ operator expanded as\footnote{To simplify the reading we avoid writing explicitly the wedge product symbol. Moreover, we introduce the notation ${\mathbb Y}_\lambda^{(n)}(x_i,\tau_i)$ to denote 
the PCO which localizes the $x_i$-integral on the curve $\tau_i \to x(\tau_i)$ parametrized by $\tau_i$. }
\begin{eqnarray}
\label{NONF}
W &=& {\rm Tr}_{\cal R} \left( 
1 +  \int_{{\cal M} \times {\mathbb T}} A^{(1)}(x)  {\mathbb Y}^{(n)}_\lambda (x,\tau)\right. \\
&& \qquad \quad \left. + \frac12 \int \! \! \! \int_{{\cal M} \times {\mathbb T}} A^{(1)}(x_1)  A^{(1)}(x_2) P\left[ 
{\mathbb Y}_\lambda^{(n)}(x_1,\tau_1) {\mathbb Y}_\lambda^{(n)}(x_2,\tau_2) 
\right] + \dots\right) \nonumber 
\end{eqnarray}
where $ {\mathbb Y}^{(n)}_\lambda$ is given in \eqref{AKACA} and localizes the integrands on a closed path $\l$, while the path-ordered product of PCOs is defined as
\begin{eqnarray}
\label{NONG}
\hspace{-0.9cm}P\left[ 
{\mathbb Y}_\lambda^{(n)}(x_1,\tau_1) {\mathbb Y}_\lambda^{(n)}(x_2,\tau_2) 
\right] &=&   \Theta(\tau_1 - \tau_2) 
{\mathbb Y}_\lambda^{(n)}(x_1, \tau_1) {\mathbb Y}_\lambda^{(n)}(x_2,\tau_2)  \\
&& \qquad + \Theta(\tau_2 - \tau_1) 
{\mathbb Y}_\lambda^{(n)}(x_1, \tau_2) {\mathbb Y}_\lambda^{(n)}(x_2, \tau_1) \nonumber 
\end{eqnarray}
We note that in (\ref{NONF}) the path ordering involves only the PCOs, since it is well-defined only for functions living on the contour. Inserting \eqref{NONG} in the $W$ expansion and performing the ${\cal M}$ integrals we are back to the usual path-ordered expansion defined in footnote 7. 

We consider the gauge variation of \eqref{NONF} under  
\be\label{variation}
\d A^{(1)} = d \omega + \, [A^{(1)}, \omega] \equiv \nabla \omega
\ee
where $\omega$ is a smooth function on the ${\cal M}$ manifold with values in the Lie algebra of ${\cal G}$. Due to the second term in this transformation, the gauge invariance of \eqref{NONF} requires cancellation of terms arising from different orders in the expansion. We are going to check gauge invariance up to cubic order in the connection.

We start discussing the variation of the linear term in \eqref{NONF}. At this order gauge invariance easily follows from the chain of identities
\begin{eqnarray}
\label{NONI}
\delta \int_{{\cal M} \times {\mathbb T}} A^{(1)}(x) {\mathbb Y}^{(n)}_\lambda(x,\tau) &=& \int_{{\cal M} \times {\mathbb T}} \left( \nabla \omega(x) {\mathbb Y}^{(n)}_\lambda(x,\tau) \right) =\int_{{\cal M} \times {\mathbb T}} \nabla  \left( \omega(x) {\mathbb Y}^{(n)}_\lambda (x,\tau) \right) \nonumber \\
&=&
 \int_{{\cal M} \times {\mathbb T}} \left[  A^{(1)}(x){\mathbb Y}^{(n)}_\lambda (x,\tau), \omega (x) \right]   
\end{eqnarray}
where in the first line we have used $\nabla {\mathbb Y}^{(n)}_\lambda= d {\mathbb Y}^{(n)}_\lambda = 0 $, being the PCO a $d$-closed, gauge singlet form. Moreover, in the second line we have neglected $d$-exact terms. 
This expression trivially vanishes when the trace is taken. 

We now move to the second order term in \eqref{NONF}. We begin by considering the contribution coming from $\delta  A^{(1)} \to d\omega$. It is explicitly given by 
\begin{eqnarray}
\label{NONJ}
 && \hspace{-0.4cm}  \delta  \,  \frac12 \int \! \! \! \int_{{\cal M} \times {\mathbb T}} A^{(1)}(x_1)  A^{(1)}(x_2) P\left[ 
{\mathbb Y}_\lambda^{(n)}(x_1,\tau_1) {\mathbb Y}_\lambda^{(n)}(x_2,\tau_2) 
\right] \nonumber \\
&& \hspace{-0.4cm} \rightarrow 
\frac12 \int \! \! \! \int_{{\cal M} \times {\mathbb T}} 
\left( d_1 \omega (x_1)  A^{(1)}(x_2) + A^{(1)}(x_1)  d_2 \omega(x_2) \right)  
P\left[ {\mathbb Y}_\lambda^{(n)}(x_1,\tau_1) {\mathbb Y}_\lambda^{(n)}(x_2,\tau_2) 
\right] \nonumber \\
&&=
\frac12 \int \! \! \! \int_{{\cal M} \times {\mathbb T}}
\left( 
\omega (x_1)  A^{(1)}(x_2) 
d_1 P\left[ {\mathbb Y}_\lambda^{(n)}(x_1,\tau_1) {\mathbb Y}_\lambda^{(n)}(x_2,\tau_2) \right] \right. \nonumber \\
&\qquad& \qquad \qquad \qquad \qquad \left.-  A^{(1)}(x_1)  \omega(x_2) 
d_2 P\left[ {\mathbb Y}_\lambda^{(n)}(x_1,\tau_1) {\mathbb Y}_\lambda^{(n)}(x_2,\tau_2) \right] 
\right) 
\end{eqnarray}
In the last step we have integrated by parts the differentials $d_1$ and $d_2$ acting on $x_1$ and $x_2$ coordinates, respectively. Now, we can use the following identities (we refer to appendix \ref{app:proof} for their proof) 
\begin{eqnarray}
\label{NONBA}
&& \hspace{-0.7cm} d_1 P \left[ {\mathbb Y}_\lambda^{(n)}(x_1,\tau_1)  {\mathbb Y}_\lambda^{(n)}(x_2,\tau_2) \right] = 
2 d\tau_1 \d(\tau_1 - \tau_2) \, \delta^{(n)}(x_1 - x_2)  \,  \bigwedge_{a=1}^n dx_1^a \wedge  {\mathbb Y}_\lambda^{(n)}(x_2,\tau_2 )  \,\nonumber \\
&& \hspace{-0.7cm} d_2 P \left[ {\mathbb Y}_\lambda^{(n)}(x_1, \tau_1)  {\mathbb Y}_\lambda^{(n)}(x_2, \tau_2) \right] = -
2  d\tau_2 \d(\tau_1 - \tau_2) \, {\mathbb Y}_\lambda^{(n)}(x_1,\tau_1)  \wedge \delta^{(n)}(x_1 - x_2)  \bigwedge_{a=1}^n dx_2^a \nonumber \\
\end{eqnarray}
where the minus sign in the second equation is due to the path ordering. We can then write \eqref{NONJ} as 
\begin{eqnarray}
\label{NONK}
&& \int \! \! \! \int_{{\cal M} \times {\mathbb T}}
\left( 
\omega (x_1)  A^{(1)}(x_2) \,  d\tau_1 \d(\tau_1 - \tau_2) \, 
 \delta^{(n)}(x_1-x_2) \bigwedge_{a=1}^n dx_1^a \wedge {\mathbb Y}_\lambda^{(n)}(x_2, \tau_2) \right.
\nonumber \\
&& 
\left. + A^{(1)}(x_1)  \omega(x_2) \, d\tau_2 \d(\tau_1 - \tau_2) \, {\mathbb Y}_\lambda^{(n)}(x_1,\tau_1)  \wedge \delta^{(n)}(x_1 - x_2)  \bigwedge_{a=1}^n dx_2^a
\right)  \nonumber \\
&& = \int_{{\cal M} \times {\mathbb T}} \left[ \omega (x),  A^{(1)}(x)  \, 
{\mathbb Y}_\lambda^{(n)}(x, \tau) \right] 
\end{eqnarray}
where in the last line we have integrated in the $(x_2, \tau_2)$ variables using the identity $\displaystyle \d(\tau_1 - \tau_2) = - \d(\tau_2 - \tau_1)$ in order to preserve the orientation of the loop. By taking the trace this term eventually vanishes. 

We now consider the contribution from the variation of the second order term in \eqref{NONF} under $\delta  A^{(1)}(x) \to [ A^{(1)}(x), \omega(x)]$,  
\begin{eqnarray}\label{NONL}
&& \hspace{-0.9cm}   \frac12 \, \d \; {\rm Tr}_{\cal R} \int \! \! \! \int_{{\cal M} \times {\mathbb T}} A^{(1)}(x_1)  A^{(1)}(x_2) P\left[ 
{\mathbb Y}_\lambda^{(n)}(x_1,\tau_1) {\mathbb Y}_\lambda^{(n)}(x_2,\tau_2) \right] \\
&& \hspace{-0.6cm}  = \frac{1}{2} \int \! \! \! \int_{{\cal M} \times {\mathbb T}}  {\rm Tr}_{\cal R}  \Big( [ A^{(1)}(x_1), \o(x_1)]  A^{(1)}(x_2) 
+ A^{(1)}(x_1)  [A^{(1)}(x_2), \o(x_2)] \Big) \nonumber \\
&\qquad& \qquad \qquad \qquad \qquad \qquad  \times P\left[ {\mathbb Y}_\lambda^{(n)}(x_1,\tau_1) {\mathbb Y}_\lambda^{(n)}(x_2,\tau_2) \right]  \nonumber \\
&& \hspace{-0.6cm}  = \int  \! \! \! \int_{{\cal M} \times {\mathbb T}}  {\rm Tr}_{\cal R}  \Big( [A^{(1)}(x_1), \o(x_1)]  A^{(1)}(x_2) \Big)
P\left[ {\mathbb Y}_\lambda^{(n)}(x_1,\tau_1) {\mathbb Y}_\lambda^{(n)}(x_2,\tau_2) \right] \nonumber
\end{eqnarray}
This term is not vanishing itself, but it is expected to compensate the variation of the cubic term in \eqref{NONF} under $\delta  A^{(1)}(x) \to d\omega(x)$. In fact, integrating by parts, the variation of the cubic term gives rise to
\begin{eqnarray}\label{NONM}
&& \hspace{-1.3cm}  \frac{1}{3!}  \, \d \, {\rm Tr}_{\cal R} \int \! \! \! \int  \! \! \! \int_{{\cal M} \times {\mathbb T}} A^{(1)}(x_1)  A^{(1)}(x_2) A^{(1)}(x_3)P\left[ 
{\mathbb Y}_\lambda^{(n)}(x_1,\tau_1) {\mathbb Y}_\lambda^{(n)}(x_2,\tau_2) {\mathbb Y}_\lambda^{(n)}(x_3,\tau_3) \right] \nonumber \\
&&\hspace{-1.2cm}  \rightarrow \frac{1}{3!}  \, \int \! \! \!\int \! \! \! \int_{\mathcal{M} \times {\mathbb T}}
{\rm Tr}_{\cal R}  \left( 
- \omega (x_1)  A^{(1)}(x_2) A^{(1)}(x_3) 
\; d_1 P\left[ {\mathbb Y}_\lambda^{(n)}(x_1,\tau_1) {\mathbb Y}_\lambda^{(n)}(x_2,\tau_2) {\mathbb Y}_\lambda^{(n)}(x_3,\tau_3)\right] \right. \nonumber \\
&\qquad& \qquad \qquad \qquad \quad \left.+ A^{(1)}(x_1)  \omega(x_2) A^{(1)}(x_3) 
\; d_2 P\left[ {\mathbb Y}_\lambda^{(n)}(x_1,\tau_1) {\mathbb Y}_\lambda^{(n)}(x_2,\tau_2) {\mathbb Y}_\lambda^{(n)}(x_3,\tau_3)\right] \right. \nonumber \\
&\qquad& \qquad \qquad \qquad \quad \left.- A^{(1)}(x_1)  A^{(1)}(x_2) \omega(x_3) 
\; d_3 P\left[ {\mathbb Y}_\lambda^{(n)}(x_1,\tau_1) {\mathbb Y}_\lambda^{(n)}(x_2,\tau_2) {\mathbb Y}_\lambda^{(n)}(x_3,\tau_3)\right]  
\right) \nonumber \\
\end{eqnarray}
As described in appendix \ref{app:proof} (see for instance eq. \eqref{DFE3}), the action of the $d_{j=1,2,3}$ differential on the path-ordered product of PCO's has the net effect to 
replace\footnote{Here we use the cycling convention $\tau_0 \equiv \tau_3$, $\tau_4 \equiv \tau_1$.} $\Theta(\tau_{j-1} -\tau_j)$ with $-d\tau_j \d(\tau_{j-1} -\tau_j)$ and correspondingly ${\mathbb Y}_\lambda^{(n)}(x_j,\tau_j)$ with $ \d^{(n)}(x_{j-1} - x_j) \left( \bigwedge_{a=1}^{n} dx_j^a \right) $; or $\Theta(\tau_{j} -\tau_{j+1})$ with $d\tau_j \d(\tau_{j} -\tau_{j+1})$ and ${\mathbb Y}_\lambda^{(n)}(x_j,\tau_j)$ with $ \d^{(n)}(x_{j} - x_{j+1}) \left( \bigwedge_{a=1}^{n} dx_j^a \right) $. In both cases the delta functions allow to perform the $d\tau_j \left( \bigwedge_{a=1}^{n} dx_j^a \right) $ integrations, so reducing \eqref{NONM} to a double integral. Moreover, having the two terms opposite sign, we can easily reconstruct a commutator $[\o(x_j) , A^{(1)}(x_j)]$ for every $j=1,2,3$. Exploiting the symmetries of the integrand under the exchange of integration variables the six terms  in each path-ordered product   give eventually the same contribution, so that we end up with    
\begin{eqnarray}
\hspace{-0.1cm} \frac{1}{3!} \; 3!  \int \! \!  \! \int_{{\cal M} \times {\mathbb T}} \! \! \! \! \! {\rm Tr}_{\cal R}   \Big( \left[ \omega ( x_1),  
A^{(1)}( x_1 ) \right]  A^{(1)} ( x_2 )  \Big)  P\left[ {\mathbb Y}_\lambda^{(n)}(x_1,\tau_1) {\mathbb Y}_\lambda^{(n)}(x_2,\tau_2) \right]  \nonumber \\
\end{eqnarray}
This expression cancels exactly the contribution in \eqref{NONL}. We have then proved gauge invariance of \eqref{NONF}, up to cubic order. However, it is an easy task to realize that the same pattern keeps repeating order by order, so ensuring gauge invariance of the complete Wilson operator. 

A similar analysis holds also in the case of non-abelian super-Wilson loops.

\subsection{Supersymmetry Invariance}

We now consider the non-abelian generalization of a super-Wilson loop, ${\cal W} = {\rm Tr}_{\cal R} P e^{\G}$ with super-holonomy $\G$ given in \eqref{AKAF}, and study its variation under supersymmetry transformations.  
For the abelian case, in section \ref{sect:susy} we have discussed conditions that can be imposed in order to have either local or global supersymmetry. In this section we briefly show that in the non-abelian case slight differences arise. Given an expansion for ${\cal W}$ in powers of the supergauge connection similar to \eqref{NONF}, we will restrict to the study of the linear term. 

We first recall that the supersymmetry variation of a non-abelian gauge superfield, obtained as usual by the action of a Lie derivative, can be expressed as
\begin{equation} \label{SVNAA}
	\delta_{\epsilon} A^{(1|0)} = \nabla \iota_\epsilon A^{(1|0)}  + \iota_\epsilon F^{(2|0)}  
\end{equation}
where $\nabla = d + [A^{(1|0)}, \cdot ]$ is the gauge covariant differential in superspace. The supersymmetry variation of the linear term in the ${\cal W}$ expansion can then be written as
\begin{equation} \label{SVNAB}
	  \int_{\mathcal{SM}} {\rm Tr}_{\cal R} \left( \delta_\epsilon A^{(1|0)}  \wedge \mathbb{Y}_\Lambda^{(n-1|m)} \right) =  \int_{\mathcal{SM}} {\rm Tr}_{\cal R} \left[ \nabla \left( \iota_\epsilon A^{(1|0)}  \wedge \mathbb{Y}_\Lambda^{(n-1|m)} \right)  + \iota_\epsilon F^{(2|0)}  \wedge \mathbb{Y}_\Lambda^{(n-1|m)} \right]
\end{equation}
where we have used $\nabla \mathbb{Y}_\Lambda^{(n-1|m)} = 0$. For closed paths the first term vanishes identically. Therefore, the only term that may affect the supersymmetry invariance of ${\cal W}$ is the second one.
Following the discussion in section \ref{sect:susy} and adapting it to the non-abelian case, this term drops out if we require
\begin{equation} \label{SVNAD}
	\iota_\epsilon F^{(2|0)}\Big|_\L = \nabla \Upsilon  
\end{equation}
As in the abelian case, neglecting $\nabla$-exact terms this condition reduces to \eqref{CAE}.

\sect{Relating Wilson Loops and Pure Spinor Vertex Operators}

The Wilson loop expectation value $ \langle {\cal W} \rangle$, which describes the motion of a superparticle along a path $\L$ in a gauge background has a stringy interpretation within the AdS/CFT correspondence \cite{Maldacena:1998im, Rey:1998ik}. In fact, being the particle excited by an open massless vertex operator at the boundary of the string worldsheet, the Wilson loop expectation value equals the string partition function on a worldsheet ending on $\L$ at the boundary\footnote{This is strictly true for Wilson loops in fundamental representation. For Wilson loops in higher-dimensional representations the dual description is in terms of D3- or D5-brane configurations \cite{Drukker:2005kx, Gomis:2006sb, Gomis:2006im}.}. In particular, in the $\a' \to 0$ limit the partition function can be computed semiclassically and leads to a prediction for $\langle {\cal W} \rangle$ at strong coupling \cite{Drukker:1999zq, Maldacena:1998im, Rey:1998ik}. 

On the other hand, in the pure spinor approach to string theory the integrated vertex operator for the massless spectrum of the open superstring reads \cite{Berkovits:2001rb}
\begin{eqnarray}
\label{PSC}
V = \int d\tau \left( A_a \Pi^a +A_\a \dot \theta^\a +  F^{ab} N_{ab} + W^\a d_\a 
\right)
\end{eqnarray}
where $d_\a$ is a  worldsheet field related to the conjugate momentum to $\theta^\a$ and $N_{ab}$ is the 
Lorentz generator in the pure spinor space. At quantum level it is invariant under the BRST transformations \cite{Berkovits:2001rb} 
\begin{eqnarray}
\label{PSD}
Q d_\a = \Pi^a (\gamma_a \lambda)_\a \ , ~~~~~~
Q N_{ab} = d^\a (\gamma_{ab})_{\a \b} \lambda^\b
\end{eqnarray}
where the nilpotency conditions $Q^2 d_\a= Q^2 N_{ab} =0$ follow from the requirement for $\l_\a$ to be a commuting pure spinor, i.e. $\lambda\gamma^a \lambda=0$.

Comparing equation \eqref{PSC} with the expression for the ten dimensional superholonomy $\G$ given in \eqref{BKC} we see that the first two terms are identical. Therefore, from the perspective of relating Wilson loops to open string worldsheets, we investigate whether it is possible to modify $\G$ in such a way to obtain an expression formally identical to the string vertex operator. Indeed, we show that this is possible by applying a $d$-deformation to the PCO \eqref{BKA3} along the lines described in section \ref{sect:susyWL}. 

To prove this statement we deform the original PCO in \eqref{BKA3} as 
\begin{eqnarray} \label{PSF}
{\mathbb Y}^{(10|16)}_\L &\to& {\mathbb Y}^{(10|16)}_\L+ d \S^{(9|16)} \\
\Sigma^{(9|16)} &=& d \tau \epsilon_{a_1 \dots a_{10}} V^{a_1} \wedge V^{a_8} N^{a_9 a_{10}} \delta^{16}(\psi) + 
d \tau \epsilon_{a_1 \dots a_{10}} V^{a_1} \wedge V^{a_9} d_\a (\gamma^{a_{10}})^{\a\b} \frac{\partial}{\partial \psi^\b}\,  \delta^{16}(\psi) \nonumber
\end{eqnarray}
where we have introduced the two-vector $N_{ab}$ and the ten dimensional spinor $d_\a$. In order to compute $d \Sigma^{(9|16)}$ we need to specify how the differential acts on these new fields. In analogy with the action of the $Q$ operator in eq. \eqref{PSD} we propose
\begin{eqnarray}
\label{PSH}
d  \,d_\a = V^a (\gamma_a \psi)_\a\,, ~~~~~
d N_{ab} = d^\a (\gamma_{ab})_{\a\b} \psi^\b -\frac12 V_a\wedge V_b
\end{eqnarray}
We note that, without imposing any pure spinor constraint, these definitions automatically satisfy the Bianchi identities $d^2 d_\a = 0$ and $d^2 N_{ab} =0$. In particular, for $N_{ab}$ this is guaranteed by the addition of the extra term $-\frac12 V_a\wedge V_b$, which is instead absent in \eqref{PSD}. 

Given these definitions we can now evaluate how the original superholonomy gets modified. Recalling that 
\begin{eqnarray}
\label{PSA}
\G \rightarrow \Gamma' = \G + \int_{\cal SM \times \mathbb{T}} F^{(2|0)} \wedge \Sigma^{(9|16)} 
\end{eqnarray}
we focus only on the new term proportional to $F^{(2|0)}$. Inserting \eqref{PSF} and the superfield strength \eqref{SYMB}, it is explicitly given by (we shortly indicate $d_\a (\gamma^{a_{10}})^{\a\b} \frac{\partial}{\partial \psi^\b} \equiv d \gamma^{a_{10}} \iota$)
\begin{eqnarray}
\label{PSM}
&& \int_{{\cal SM} \times \mathbb{T}} 
\left( V^a\wedge V^b F_{ab} + (\psi \gamma_a W) V^a\right) \wedge 
\left( 
 d \tau \epsilon_{a_1 \dots a_{10}} V^{a_1} \wedge V^{a_8} N^{a_9 a_{10}} \delta^{16}(\psi)  \right.
 \nonumber \\ 
 && \hspace{8cm}+ \left. 
d \tau \epsilon_{a_1 \dots a_{10}} V^{a_1} \wedge V^{a_9} d \gamma^{a_{10}} \iota \, \delta^{16}(\psi) 
\right)  \nonumber \\
&& = \int_{{\cal SM} \times \mathbb{T}}  \Big( d \tau V^a\wedge V^b F_{ab}  \wedge \epsilon_{a_1 \dots a_{10}} V^{a_1} \wedge V^{a_8} N^{a_9 a_{10}} \delta^{16}(\psi) \nonumber \\
&& \hspace{6cm} + \, d \tau  (\psi \gamma_a W) V^a \wedge \epsilon_{a_1 \dots a_{10}} V^{a_1} \wedge V^{a_9} d \gamma^{a_{10}} \iota \, \delta^{16}(\psi) \Big) \nonumber \\
\end{eqnarray}
In the first term we simply antisymmetrize the vielbeins to obtain a desidered term proportional to $F^{ab} N_{ab}$ times a factorized volume form $V^{10} \delta^{16}(\psi)$. In the second term we first integrate by parts $\iota$ on the $\psi$ spinor and, after a bit of algebra, we produce a contribution proportional to $W^\a d_\a$ times a factorized volume. In total, summing the two terms we obtain 
\begin{eqnarray}
\label{PSN}
\int_{{\cal SM} \times {\mathbb T}}  ( F^{ab} N_{ab} + W^\a d_\a) \, d \tau \, V^{10} \delta^{(16)}(\psi)
\end{eqnarray}

We can now project the integrand onto the Wilson path by performing the integrations on the supermanifold coordinates.  The obtained contributions, when added to the original $\G$ as in \eqref{PSA}, reproduce the pure spinor vertex operator \eqref{PSC} written as a supermanifold integral.

\sect{Conclusions}

We have constructed super-Wilson operators in terms of integral forms describing the immersion of the supercontour in a supermanifold. In such a formulation the corresponding superholonomy is written as an integral over the entire supermanifold and the invariance of the operator under superdiffeomorphisms becomes manifest. As a by-product, we obtain an alternative description also of the ordinary Wilson loops, which can be obtained from the supersymmetric one by setting all the spinorial coordinates to zero. We have reformulated kappa-symmetry in this language and studied the Killing spinor equations associated to supersymmetry invariance. 

We have highlighted the role of the $d$-cohomology in the construction of the Picture Changing Operators (PCO). Different PCOs corresponding to different supercontours are all comohological equivalent. Nevertheless, they may preserve a different amount of supersymmetry and more generally they exhibit a different spectrum of symmetries. In particular, it follows that by adding $d$-exact terms we can tune the BPS degree of a Wilson operator and we can easily relate two operators which preserve a different fraction of supersymmetries. As a remarkable example, we have shown how the BPS Wilson-Maldacena loop of $N=4$ SYM theory can be obtained from the non-BPS one by the addition of a suitable $d$-exact term to the ordinary non-supersymmetric PCO.

It would be interesting to generalize our formulation at quantum level to compute perturbative corrections to Wilson loops. In particular, it would be nice to understand which is the effect of the {\em $d$-varying symmetries} mechanism, in a frame where $d$-exact terms could be treated as perturbations.  More ambitiously, it could be interesting to understand how to reformulate localization in a geometrical framework and exploit our expression for the Wilson loop to compute its vacuum expectation value exactly.

Finally, as emphasised in the paper, this formalism allows for a straightforward generalization to curved supermanifolds, hence leading to Wilson operators defined in a supergravity framework \cite{Castellani:2014goa,DHoker:2019clx}. This geometrical setting might be also applied to Wilson loops in different dimensions, for example to the well known bosonic BPS Wilson loops in three dimensional Chern-Simons-matter theories \cite{Drukker:2008zx,Chen:2008bp,Bianchi:2014laa,Drukker:2019bev}.
This formalism is ready also for describing higher dimensional (BPS) defects, as it only requires to choose the suitable PCO dual to the immersion of the defining hypersurface. 

\section*{Acknowledgement}

We enjoyed discussions with Leonardo Castellani, Luca Griguolo and Domenico Seminara. This work has been partially supported by Universit\`a del Piemonte Orientale research funds, by Italian Ministero dell'Universit\`a e della Ricerca (MIUR), and by Istituto Nazionale di Fisica Nucleare (INFN) through the ``FieLds And Gravity'' (FLAG) and ``Gauge theories, Strings, Supergravity'' (GSS) research projects.

\newpage
\appendix
\sect{Superspace conventions}
\label{app1}

In this appendix we collect the conventions on supermanifolds that we have used in the main text. To be definite we focus on the ${\cal N}=1$ superspace in ten dimensions
described by coordinates $z^M = (x^\mu, \theta^\a)$, with $\mu = 0, \dots , 9$ and $\a = 1, \dots , 16$. Here $\theta^\a$ are Majorana-Weyl spinors.  

Introducing ten dimensional $16 \times 16$ gamma matrices $\g^a_{\a\b}$ \footnote{Similarly, we can use matrices $(\gamma^{a})^{ \a\b}$ 
which have the same numerical values as $\gamma^a_{\a\b}$.}, supercharges and supercovariant derivatives are defined as
\begin{eqnarray}
\label{supercharges}
Q_\a = \partial_\a + \theta^\b \g^a_{\a\b} \, \partial_a \qquad , \qquad D_\a = \partial_\a - \theta^\b \g^a_{\a\b} \, \partial_a
\end{eqnarray}
with $Q_\a = D_\a + 2\theta^\b \g^a_{\a\b} \, \partial_a$. They satisfy 
\begin{eqnarray}
\{ Q_\a , Q_\b \} = 2 \g^a \partial_a \quad , \quad \{ D_\a , D_\b \} = -2 \g^a \partial_a \quad , \quad \{Q_\a , D_\b \} = 0
\end{eqnarray}
Flat supervielbeins are defined as
\begin{eqnarray}
\label{supervielbeins}
V^a \equiv e_M^a dz^M = dx^a + \theta^\a \gamma^a_{\a\b} d\theta^\b \quad , \quad \psi^\a \equiv e^\a_M dz^M = d\theta^\a
\end{eqnarray}
and satisfy the Maurer-Cartan equations 
\begin{eqnarray}
\label{genEA}
d V^a = \psi \gamma^a \psi \quad , \quad d \psi^\a=0
\end{eqnarray}
 
Introducing the ordinary super-differential basis $(dx^a, d\theta^\a)$ through the defining identities $\langle \partial_a, dx^b \rangle = \d_a^b$ and $\langle \partial_\a, d\theta^\b \rangle = \d_\a^\b$, the differential operator $d$ is given by
\begin{eqnarray}
d \equiv dx^a \partial_a + d\theta^\a \partial_\a = V^a \partial_a + \psi^\a D_\a
\end{eqnarray} 

\vskip 10pt
\noindent
{\bf Supersymmetry transformations.} In the present framework, a supersymmetry transformation is a superdiffeomorphism whose action on differential forms is represented by a Lie derivative along the vector field $\e = \e^\a D_\a + 2\e \g^a \theta  \partial_a \equiv \e^\a Q_\a$,
\begin{eqnarray}\label{Lieder}
{\cal L}_\epsilon = \iota_\epsilon d + d \iota_\epsilon  
\end{eqnarray}  
where in general $\iota_v$ is the contraction operator defined on a $p$--form as $\iota_v \omega(v_1, \dots, v_{p-1}) = \omega (v_1, \dots, v_{p-1},v)$. Using $\iota_\epsilon \theta^\a = \iota_\epsilon x^a  =0$, it then follows that
\begin{eqnarray}
\label{AKAMA}
{\cal L}_\epsilon \theta^\a &=& ( \iota_\epsilon d + d\iota_\epsilon )\theta^\a = \iota_\e d\theta^\a =  \epsilon^\a \nonumber \\ 
{\cal L}_\epsilon x^a &=& ( \iota_\epsilon d +  d\iota_\epsilon ) x^a = \iota_\epsilon dx^a = \epsilon \gamma^a \theta
\end{eqnarray}
which are the ordinary supersymmetry transformations of the superspace coordinates. In particular, these defining identities follow
\begin{eqnarray}
\label{AKAK}
\iota_\epsilon V^a =  2 \epsilon\gamma^a \theta \,, ~~~~~       \iota_\epsilon \psi^\a = \epsilon^\a
\end{eqnarray}
As a consistency check of our conventions we find that $(V^a, \psi^\a)$ are invariant under supersymmetry as a consequence of identities \eqref{genEA} and 
\eqref{AKAMA}
\begin{eqnarray}
\label{AKAM}
{\cal L}_\epsilon \psi^\a &=&  \iota_\epsilon d \psi^\a  + d \iota_\epsilon \psi^\a = 0\,, ~~~~~\nonumber \\
{\cal L}_\epsilon V^a &=& \iota_\epsilon d V^a + d \iota_\epsilon V^a   = 2 \epsilon \gamma^a \psi 
+  d ( 2 \epsilon \gamma^a \theta) =0
\end{eqnarray}

The supersymmetry variation of a scalar function $\Phi$ on the supermanifold is defined as
\begin{eqnarray}
\label{diaB}
\delta_\epsilon \Phi &=& {\cal L}_\epsilon \Phi = d \iota_\epsilon \Phi + \iota_\epsilon d \Phi = 
\iota_\epsilon (V^a \partial_a \Phi + \psi^\a D_\a \Phi) \nonumber \\
&=&  2 \epsilon \gamma^a \theta \partial_a \Phi + \epsilon^\a D_\a \Phi = \epsilon^\a Q_\a \Phi
\end{eqnarray}
where $Q_\a$ is the supersymmetry generator introduced in eq. \eqref{supercharges}.

Similarly, the supersymmetry variation of a superconnection $A^{(1|0)}$ reads
\begin{eqnarray}
\label{AKAN}
{\cal L}_\epsilon A^{(1|0)} =  d \left( \iota_\epsilon  A^{(1|0)} \right) + \iota_\epsilon F^{(2|0)}
\end{eqnarray}
where the first term is a gauge transformation of $A^{(1|0)}$ and in the second term $F^{(2|0)}$ is the superfield strength defined as $F^{(2|0)} = d A^{(1|0)} + [A^{(1|0)} , A^{(1|0)} ]$. 

\vskip 10pt
\noindent
{\bf Kappa-symmetry transformations.} A kappa-symmetry transformation is generated by a vector $\widetilde \kappa \equiv \kappa^\a D_\a$ which differs from the supersymmetry generator by the simple replacement 
\begin{eqnarray}
\label{kappaAA}
\epsilon = \epsilon^\a Q_\a \longrightarrow  \widetilde \kappa = \kappa^\a D_\a = \kappa^\a Q_\a - 2 \kappa\gamma^a \theta \, \partial_a 
\end{eqnarray}
In particular, it follows that a kappa-symmetry transformation can be formally written as the action of the Lie derivative ${\cal L}_{\widetilde \kappa}  = \{ \iota_{\widetilde \k}, d \}$ where 
\begin{eqnarray}
\label{contractions}
 \iota_{\widetilde \k} = \iota_\k - 2 \k\gamma^a \theta \iota_a 
\end{eqnarray}
When $\iota_{\widetilde \k} $ acts on the flat supervielbeins $(V^a, \psi^\a)$, using  $\iota_a V^b = \delta^b_a$ we find 
\begin{eqnarray}
\label{kappaAB}
\iota_{\widetilde \k} \psi^\a = \k^\a\,, 
~~~~~
\iota_{\widetilde \k} V^a = \iota_\k V^a - (2 \k \gamma^b \theta) \iota_b V^a = 0 
\end{eqnarray}
Therefore, kappa-symmetry transformations of the superspace coordinates and the basic one-forms read
\begin{eqnarray}
	\label{KSJ} \delta_{\widetilde \k} \theta^\alpha &=& {\cal L}_{\widetilde \k} \theta^\a =  \k^\beta \iota_\beta \psi^\alpha = \k^\alpha  \\
	\label{KSK} \delta_{\widetilde \k} x^a &=& {\cal L}_{\widetilde \k} x^a = \k^\alpha \iota_\alpha \left( dx^a - \theta \gamma^a \psi \right) + \k^\alpha \iota_\alpha \theta \gamma^a \psi = \k \gamma^a \theta  \\
	\label{KSL} \delta_{\widetilde \k} \psi^\a &=& {\cal L}_{\widetilde \k} \psi^\a = d (\iota_{\widetilde \k} \psi^\a)  = d \k^\a   \\
	\label{KSM} \delta_{\widetilde \k} V^a &=& {\cal L}_{\widetilde \k} V^a = \iota_{\widetilde \k} d V^a =  \iota_{\widetilde \k} ( \psi \gamma^a \psi) = 2 \k \gamma^a \psi
\end{eqnarray}
In the case of rigid symmetry, $d \k^\a =0$. 

It is interesting to note that the replacement of the $Q$ with the $D$ generators in \eqref{kappaAA} leads to the following dual situation between supersymmetry and kappa-symmetry transformations of the bosonic supervielbein
\be
\hbox{$\left\{ \begin{array}{cc}
\iota_\e V^a \neq 0 \\
{\cal L}_\e V^a = 0 
\end{array} \right.$}  \qquad \qquad \qquad \hbox{$\left\{ \begin{array}{cc}
\iota_{\widetilde \k} V^a = 0 \\
{\cal L}_{\widetilde \k} V^a \neq 0 
\end{array} \right.$}
\ee

Applied to a generic superfield $\Phi$, a kappa-symmetry transformation reads
\begin{eqnarray}
\label{kappaAD}
\delta_{\widetilde \k} \Phi = {\cal L}_{\widetilde \k}  \Phi = 
(\iota_{\widetilde \k} d + d 
\iota_{\widetilde \k})\Phi = \iota_{\widetilde \k} (V^a \partial_a \Phi + \psi^\a D_\a \Phi) = 
\k^\a D_\a \Phi  
\end{eqnarray}
and can be correctly obtained from a supersymmetry transformation (\ref{diaB}) by replacing the $Q_\a$ generator with $D_\a$. 

Parametrizing the ${\tilde \k}$ generator as 
\begin{equation} 
{\tilde \k} = ( \gamma^a )^{\alpha \beta} \mathcal{L}_a K_\beta \, D_\a
\end{equation}
with $K_\b$ a $0$-form and $\mathcal{L}_a$ the infinitesimal translation operator, we shift $K_\beta$ as
\begin{equation}\label{KSH}
	K_\beta  \rightarrow  K'_\beta = K_\beta +   \mathcal{L}_b \, {\cal K}^\gamma  ( \gamma^b )_{\gamma \beta}  
\end{equation}
Consequently, the $\k^\a$ parameter transforms as
\be
 \k^\alpha \rightarrow  \k^\a + ( \gamma^a \gamma^b)^\alpha_\gamma \, \mathcal{L}_a \mathcal{L}_b \, {\cal K}^\gamma  \sim \k^\a + \frac{1}{2} \partial^2 
 {\cal K}^{\alpha} 
\ee
where we have exploited $[{\cal L}_a , {\cal L}_b ] = 0$. 
If ${\cal K}$ is an \emph{harmonic} function, then transformation \eqref{KSH} is a symmetry of the kappa-symmetry parameter, i.e. $K_\b$ and $ K'_\b$ give rise to the same kappa-symmetry transformation. This degeneracy can be used to halve the number of independent components of the $K$ spinor.

\sect{An alternative expression for the ordinary Wilson loop}

In this appendix we show that we can re-express formula \eqref{WLF} for the bosonic Wilson loop as
\begin{eqnarray}
\int_{\cal M} A^{(1)}\wedge {\mathbb Y}^{(n-1)}_\l =  \int_{\cal M} \text{Vol}^{(n)} \prod_{a=1}^{n-1} \delta \left( \phi_a \right) \ ,
\end{eqnarray}
where $\text{Vol}^{(n)}$ is the volume form on $\cal M$, given by $\text{Vol}^{(n)} = A^{(1)} \wedge \prod_{a=1}^{n-1} d \phi_a$.

This formula can be proved as follows. Given the Poincar\'e dual $\displaystyle {\mathbb Y}^{(n-1)}_\l$ in eq. \eqref{WLC} that defines the immersion of the $\lambda$ curve, we choose a (local) basis of vectors $\displaystyle \left\lbrace X^a \right\rbrace_{a=1}^{n-1} $ of $\lambda^\perp$ normalised by $\iota_{a} d \phi^b \equiv X^a \left( \phi_b \right) = \delta^a_b $.
Then, it is easy to prove that
\begin{eqnarray}
	\prod_{a=1}^{n-1} \iota_{a} \text{Vol}^{(n)} = A^{(1)} 
\end{eqnarray}
and we can write the following chain of identities
\begin{equation}
	\int_{\cal M} A^{(1)} \wedge \mathbb{Y}^{(n-1)}_\l  = \int_{\cal M} A^{(1)} \prod_{a=1}^{n-1} d \phi_a \delta \left( \phi_a \right) = \int_{\cal M} \left[ \prod_{a=1}^{n-1} d \phi_a \delta \left(\phi_a \right) \iota_a \right] \text{Vol}^{(n)} = $$ $$ = \int_{\cal M} \text{Vol}^{(n)} \prod_{a=1}^{n-1} \delta \left( \phi_a \right) 
\end{equation}
where in the last equality we have used the Leibnitz rule and
\begin{equation}
	\iota_{X_{a_1}} \left[ d \phi_{a_1} \wedge \ldots \wedge d \phi_{a_{n-1}} \iota_{X_{a_{2}}} \ldots \iota_{X_{a_{n-1}}} \text{Vol}^{(n)} \right] = 0
\end{equation}
since the form inside the brackets is an $(n+1)$-form in an $n$-dimensional manifold.

\sect{Proof of identity \eqref{NONBA}}\label{app:proof}

Here we want to compute the following expression 
\begin{eqnarray}
	\label{DFB} 
d_j P \left[ {\mathbb Y}_\lambda^{(n)}(x_1,\tau_1) \wedge  \ldots   \wedge {\mathbb Y}_\lambda^{(n)}(x_M, \tau_M ) \right]
\end{eqnarray}
where $\displaystyle d_j \equiv dx_j^a \tfrac{\partial}{\partial x_j^a} + d \tau_j \tfrac{\partial}{\partial \tau_j}$, $j = 1 , \ldots , M $, denotes the exterior derivative w.r.t. the set of coordinates $(x_j^a, \tau_j)$ acting on the path-ordered wedge product of $M$ PCO's of the form \eqref{AKACA}. This expression is required to prove gauge invariance of the non-abelian Wilson operator, as discussed in section \ref{sect:gaugeinvariance}. 

All the PCO's localize on the same contour, but parametrized by different parameters $\tau_i \in \mathbb{T}$. The definition of the corresponding path-ordering is\footnote{In order to avoid cluttering, in what follows we will neglect the wedge symbol.}
\begin{eqnarray}
	\label{DFC} 
&&P \left[ {\mathbb Y}_\lambda^{(n)}(x_1,\tau_1) \ldots   {\mathbb Y}_\lambda^{(n)}(x_M,\tau_M ) \right]  \\
&& = \sum_{\sigma \in \Sigma} \Theta \left( \sigma (\tau_1) - \sigma (\tau_2) \right) \Theta \left( \sigma (\tau_2) - \sigma (\tau_3) \right) \ldots \Theta \left( \sigma (\tau_{M-1}) - \sigma (\tau_M) \right) \nonumber \\
&& \qquad \quad  \times {\mathbb Y}_\lambda^{(n)}(x_1,\sigma (\tau_1) )   \ldots   {\mathbb Y}_\lambda^{(n)}(x_M, \sigma (\tau_M) ) \nonumber
\end{eqnarray}
where $\Sigma$ denotes all the possible $M!$ permutations of $\{ \tau_1, \dots , \tau_M \}$.

As a warming-up we compute \eqref{DFB} for $M=2$. From the previous definitions, recalling that $d \mathbb{Y}_\lambda^{(n)} = 0$, we can write
\begin{eqnarray}\label{NONBD}
&& \hspace{-0.4cm}d_1   P\left[ {\mathbb Y}_\lambda^{(n)}(x_1, \tau_1) \, {\mathbb Y}_\lambda^{(n)}(x_2,\tau_2) \right]   =  d_1 \Theta(\tau_1 - \tau_2) \; {\mathbb Y}_\lambda^{(n)}(x_1, \tau_1) \, {\mathbb Y}_\lambda^{(n)}(x_2,\tau_2) + \, \tau_1   \leftrightarrow \tau_2  \, \nonumber \\
&= & d \tau_1 \delta( \tau_1 \! - \! \tau_2 ) \delta^{(n)} ( x_1 \! - \! x ( \tau_1  )  ) ( dx_1 \!- \! \dot{x} ( \tau_1) d \tau_1 )^n   \delta^{(n)} ( x_2  \! - \! x ( \tau_2 ) ) ( dx_2 \! - \! \dot{x}( \tau_2)d \tau_2 )^n+ \tau_1 \leftrightarrow \tau_2   \nonumber \\
&=& 2 d \tau_1 \,  \delta ( \tau_1 - \tau_2 ) \, \delta^{(n)} ( x_1 - x_2 ) \, d^n x_1 \, \mathbb{Y}_\lambda^{(n)}( x_2,\tau_2) \nonumber 
\end{eqnarray}
where in the last step the product of all the delta functions has been used to generate $\delta^{(n)} \left( x_1 - x_2 \right)$. 
Similarly, it is easy to realize that applying the $d_2$ differential we end up with 
\begin{eqnarray}\label{NONBD2}
&& \hspace{-0.4cm}d_2   P\left[ {\mathbb Y}_\lambda^{(n)}(x_1, \tau_1) \, {\mathbb Y}_\lambda^{(n)}(x_2,\tau_2) \right]   
= -2 d \tau_2 \,  \delta ( \tau_1 - \tau_2 ) \, \mathbb{Y}_\lambda^{(n)}( x_1,\tau_1) \, \delta^{(n)} ( x_1 - x_2 ) \, d^n x_2  \nonumber 
\end{eqnarray}
where the minus sign comes from applying $d_2$ to $\Theta(\tau_1 - \tau_2)$. This is indeed the sign that turns out to be crucial for producing eventually the integral of a commutator (see eq. \eqref{NONK}).

We now generalize the calculation to the product of $M$ PCO's. For the sake of clarity, we focus on a single term of \eqref{DFC}, namely 
\begin{equation}
	\Theta \left( \tau_1 - \tau_2 \right) \Theta \left( \tau_2 - \tau_3 \right) \ldots \Theta \left( \tau_{M-1} - \tau_M \right) {\mathbb Y}_\lambda^{(n)}(x_1, \tau_1)  \ldots   {\mathbb Y}_\lambda^{(n)}(x_M,\tau_M )
\end{equation}
and first consider applying $d_1$. Since $d_1$ acts on a single theta function, we obtain the following chain of identities
\begin{eqnarray}\label{DFE}
&& \hspace{-0.8cm} d_1 \left[ \Theta \left( \tau_1 - \tau_2 \right) \Theta \left( \tau_2 - \tau_3 \right) \ldots \Theta \left( \tau_{M-1} - \tau_M \right) {\mathbb Y}_\lambda^{(n)}(x_1,\tau_1 )  \ldots   {\mathbb Y}_\lambda^{(n)}(x_M,\tau_M)\right] \nonumber \\
&=& d\tau_1 \, \delta( \tau_1 - \tau_2 ) \, \, \Theta ( \tau_2 - \tau_3 ) \ldots \Theta ( \tau_{N-1} - \tau_N ) {\mathbb Y}_\lambda^{(n)}(x_1,\tau_1)   \ldots  {\mathbb Y}_\lambda^{(n)}(x_M ,\tau_M ) \nonumber \\
& =& d\tau_1 \delta ( \tau_1 - \tau_2 ) \, \, \Theta ( \tau_2 - \tau_3 ) \ldots \Theta ( \tau_{M-1} - \tau_M )  \\
&\qquad & \qquad \times  \, \left( \d^{(n)}(x_1 - x_2) \, \bigwedge_{a=1}^{n} dx_1^a \right) \wedge {\mathbb Y}_\lambda^{(n)}(x_2, \tau_2 )  \wedge \ldots \wedge {\mathbb Y}_\lambda^{(n)}(x_M,\tau_M )  \nonumber
\end{eqnarray}
If we now apply $d_2$ we have to take into account that this time the differential acts on two different theta functions. Therefore, in this case we obtain 
\begin{eqnarray}\label{DFE2}
&& \hspace{-0.8cm} d_2 \left[ \Theta \left( \tau_1 - \tau_2 \right) \Theta \left( \tau_2 - \tau_3 \right) \ldots \Theta \left( \tau_{M-1} - \tau_M \right) {\mathbb Y}_\lambda^{(n)}(x_1,\tau_1  )  \ldots   {\mathbb Y}_\lambda^{(n)}(x_M,\tau_M )\right]  \\
&=& - \, d\tau_2 \, \delta ( \tau_1 - \tau_2 ) \, \Theta ( \tau_2 - \tau_3 ) \Theta ( \tau_3 - \tau_4 )\ldots \Theta ( \tau_{M-1} - \tau_M )   \nonumber \\
&&   \times \,  {\mathbb Y}_\lambda^{(n)}(x_1,\tau_1 )  \wedge \left( \d^{(n)}(x_1 - x_2) \, \bigwedge_{a=1}^{n} dx_2^a \right)  \wedge {\mathbb Y}_\lambda^{(n)}(x_3 ,\tau_3 )  \wedge \ldots \wedge {\mathbb Y}_\lambda^{(n)}(x_M,\tau_M )  \nonumber \\
&~& + \,  \Theta ( \tau_1 - \tau_2 ) \, d\tau_2 \, \delta ( \tau_2 - \tau_3 ) \,  \Theta ( \tau_3 - \tau_4 ) \ldots \Theta ( \tau_{M-1} - \tau_M )   \nonumber \\
&&   \times\,  {\mathbb Y}_\lambda^{(n)}(x_1,\tau_1 ) \wedge \left( \d^{(n)}(x_2 - x_3) \, \bigwedge_{a=1}^{n} dx_2^a \right) \wedge {\mathbb Y}_\lambda^{(n)}(x_3,\tau_3 )    \wedge \ldots \wedge {\mathbb Y}_\lambda^{(n)}(x_M,\tau_M )  \nonumber
\end{eqnarray}
We see that the first term is exactly {\em minus} the term in \eqref{DFE}, whereas the second term will coincide with {\em minus} one of the two terms which arise when we apply $d_3$. This pattern repeats itself for any other differential acting on intermediate theta functions.  The $d_j$ differential will produce two terms
\begin{eqnarray}\label{DFE3}
&& \hspace{-0.8cm} d_j \left[ \Theta \left( \tau_1 - \tau_2 \right) \Theta \left( \tau_2 - \tau_3 \right) \ldots \Theta \left( \tau_{M-1} - \tau_M \right) {\mathbb Y}_\lambda^{(n)}(x_1,\tau_1  )  \ldots   {\mathbb Y}_\lambda^{(n)}(x_M, \tau_M  )\right]   \\
&=& - \,  \Theta ( \tau_1 - \tau_2 ) \dots d\tau_j \, \delta ( \tau_{j-1}- \tau_j ) \,\Theta ( \tau_j - \tau_{j+1} )\ldots \Theta ( \tau_{M-1} - \tau_M )   \nonumber \\
&&  \times  \, {\mathbb Y}_\lambda^{(n)}(x_1,\tau_1 )  \ldots \left( \d^{(n)}(x_{j-1} - x_j) \, \bigwedge_{a=1}^{n} dx_j^a \right)  {\mathbb Y}_\lambda^{(n)}(x_{j+1},\tau_{j+1} ) \ldots   {\mathbb Y}_\lambda^{(n)}(x_M,\tau_M )  \nonumber \\
&~& + \,  \Theta ( \tau_1 - \tau_2 ) \ldots  d\tau_j \, \delta ( \tau_j - \tau_{j+1} ) \,  \Theta ( \tau_{j+1} - \tau_{j+2}) \ldots \Theta ( \tau_{M-1} - \tau_M )  \nonumber \\
&&  \times  \, {\mathbb Y}_\lambda^{(n)}(x_1,\tau_1 ) \ldots \left( \d^{(n)}(x_{j} - x_{j+1}) \, \bigwedge_{a=1}^{n} dx_j^a \right)  {\mathbb Y}_\lambda^{(n)}(x_{j+1},\tau_{j+1} ) \ldots   {\mathbb Y}_\lambda^{(n)}(x_M,\tau_M )  \nonumber
\end{eqnarray}
the first one being opposite in sign to a term coming from $d_{j-1}$ and the second one opposite to a term from the application of $d_{j+1}$.
 
 The same pattern holds for any other term of \eqref{DFC} when we consider the contributions coming from the application of $d_{\sigma (\tau_{i-1})}$, $d_{\sigma (\tau_i)}$ and $d_{\sigma (\tau_{i+1})}$ on the product $\Theta \left( \sigma (\tau_{i-1}) - \sigma (\tau_{i}) \right)\Theta \left( \sigma (\tau_i) - \sigma (\tau_{i+1}) \right)$). Precisely, the $d_{\sigma (\tau_i)}$ derivative produces two terms which come in pair with opposite signs with one term from $d_{\sigma (\tau_{i-1})}$ and one from $d_{\sigma (\tau_{i+1})}$. As we discuss in section \ref{sect:gaugeinvariance}, these signs are crucial for reconstructing commutators and ensure cancellation in the gauge variation of the Wilson loop. 
 
It is important to observe that an identical proof works also in the case of a supermanifold $\mathcal{SM}$, i.e. for products of super-Poincar\'e duals  of the form ${\mathbb Y}^{(n|0)} \wedge {\mathbb Y}^{(0|m)}$(see eq. \eqref{AKAEgeneral}) localizing super-integrals on supercontours.

\newpage

\end{document}